\documentclass[sigconf,screen,hyperref={colorlinks = true,linkcolor = blue}]{acmart}

\AtBeginDocument{%
  }
\setcopyright{none} 
\copyrightyear{2025}
\acmYear{2025}
\acmDOI{XXXXXXX.XXXXXXX}

\acmConference[Preprint]{Preprint}{2025}{arXiv}

\acmISBN{} 
\acmDOI{}  

\renewcommand\footnotetextcopyrightpermission[1]{} 
\usepackage{fontawesome}
\usepackage{txfonts}
\usepackage{utfsym}
\usepackage{savesym}
\savesymbol{Asterisk}
\usepackage{bbding}
\restoresymbol{BBD}{Asterisk}
\usepackage{mathabx}
\usepackage{packages/packages}
\usepackage{packages/macros}
\usepackage{titlesec}
\titlespacing*{\section}{0pt}{1.5ex}{1ex}
\titlespacing*{\subsection}{0pt}{1ex}{0.5ex}
\begin{document}
\title[\ours]{\ours: Systematic Performance Optimization \\for Retrieval-Augmented Generation Serving}

\DeclareRobustCommand{\robustemail}[1]{%
  \href{mailto:#1}{\texttt{#1}}%
}

\author{Wenqi Jiang$^{\footnotesize{\bullet}}$}
\affiliation{%
  \institution{ETH Zurich}
\country{\small{\href{mailto:wenqi.jiang@inf.ethz.ch}{\texttt{wenqi.jiang@inf.ethz.ch}}}}
}

\author{Suvinay Subramanian}
\affiliation{%
  \institution{Google}
\country{\small{\href{mailto:suvinay@google.com}{\texttt{suvinay@google.com}}}}
}

\author{Cat Graves}
\affiliation{%
  \institution{Google DeepMind}
\country{\small{\href{mailto:cgraves@google.com}{\texttt{cgraves@google.com}}}}
}
\author{Gustavo Alonso}
\affiliation{%
  \institution{ETH Zurich}
\country{\small{\href{mailto:alonso@inf.ethz.ch}{\texttt{alonso@inf.ethz.ch}}}}
}

\author{Amir Yazdanbakhsh$^\vardiamondsuit$}
\affiliation{%
  \institution{Google DeepMind}
\country{\small{\href{mailto:ayazdan@google.com}{\texttt{ayazdan@google.com}}}}
}

\author{Vidushi Dadu$^\vardiamondsuit$}
\affiliation{%
  \institution{Google}
\country{\small{\href{mailto:vidushid@google.com}{\texttt{vidushid@google.com}}}}
}

\thanks{$^{\tiny{\bullet}}$ Work done while at Google\\$^\vardiamondsuit$ Equal advising}

\renewcommand{\shortauthors}{Wenqi Jiang et al.} 
\renewcommand{\shorttitle}{\ours: Systematic Performance Optimization for Retrieval-Augmented Generation Serving} 

\sloppy
\begin{abstract}
Retrieval-augmented generation (RAG), which combines large language models (LLMs) with retrievals from external knowledge databases, is emerging as a popular approach for reliable LLM serving. 
However, efficient RAG serving remains an open challenge due to the rapid emergence of many RAG variants and the substantial differences in workload characteristics across them.
In this paper, we make three fundamental contributions to advancing RAG serving.
First, we introduce \ragdescemph, a structured abstraction that captures the wide range of RAG algorithms, serving as a foundation for performance optimization.
Second, we analyze several representative RAG workloads with distinct \ragdesc, revealing significant performance variability across these workloads.
Third, to address this variability and meet diverse performance requirements, we propose \oursemph (\underline{\textbf{R}}etrieval-\underline{\textbf{A}}ugmented \underline{\textbf{G}}eneration \underline{\textbf{O}}ptimizer), a system optimization framework for efficient RAG serving. 
Our evaluation shows that \ours achieves up to a 2$\times$ increase in QPS per chip and a 55\% reduction in time-to-first-token latency compared to RAG systems built on LLM-system extensions.

\end{abstract}

\maketitle

\section{Introduction}
\label{sec:intro}
\begin{figure}[t]
  \centering
  \includegraphics[width=1.0\linewidth]{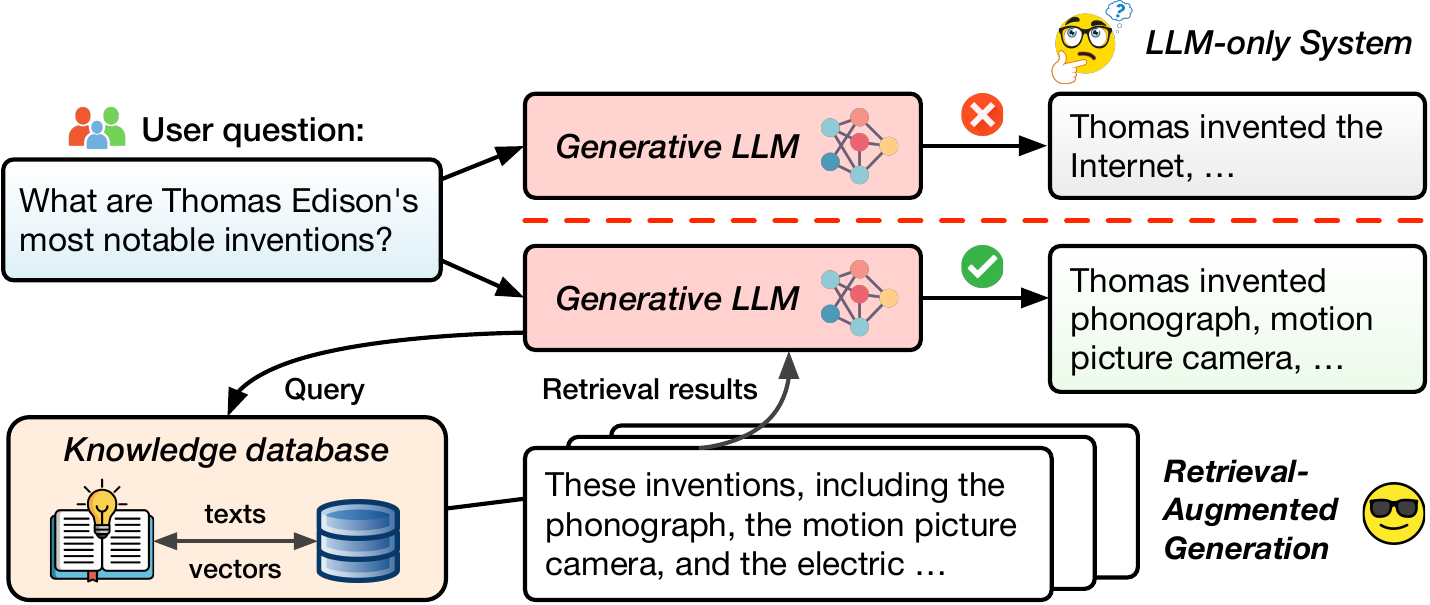}
  \vspace{-1.5em}
  \caption{LLM-only system (top) versus RAG (bottom).}
  \vspace{-1em}
  \label{fig:rag-example}
\end{figure}
The rapid adoption of Large Language Models (LLMs) across diverse applications\,---\,such as question answering~\cite{brown2020language, chowdhery2022palm, dubey2024llama}, code generation~\cite{li2022competition, liu2024your, poesia2022synchromesh}, and scientific discoveries~\cite{beltagy2019scibert, taylor2022galactica, lee2020biobert}\,---\,showcases their profound impact on automating knowledge-based tasks.
Despite these capabilities, LLM systems, when deployed in isolation (aka LLM-only systems), face substantial challenges, such as data staleness~\cite{lazaridou2021mind, lewis2020retrieval}, a propensity to hallucinate (generating factually incorrect or nonsensical information), and limited, often rigid model knowledge~\cite{lewis2020retrieval,li2023dark, ji2023survey}.
These challenges hinder the reliability and adaptability of LLM-only systems, especially in applications that demand high factual accuracy~\cite{izacard2020leveraging, fan2019eli5, kasneci2023chatgpt, thirunavukarasu2023large}.

Retrieval-Augmented Generation (RAG) has emerged as one powerful solution to address the common pitfalls of LLM-only systems for knowledge-intensive tasks~\cite{borgeaud2022improving, shao2024scaling, wang2023instructretro, lazaridou2021mind, lewis2020pre}.
By retrieving information from external databases and appending it to prompts (\Cref{fig:rag-example}), RAG enhances the credibility, timeliness, and contextually rich nature of LLM-generated responses.
Leveraging the generative prowess of LLMs alongside external knowledge sources, RAG not only achieves comparable quality with smaller models~\cite{borgeaud2022improving, shao2024scaling, wang2023instructretro} but also simplifies the process of updating knowledge, mitigating the extent of additional model training, which is often prohibitively expensive~\cite{lazaridou2021mind, lewis2020pre}.
These advantages have established RAG as the industry standard for knowledge-intensive applications, with notable examples including Google's REALM~\cite{guu2020realm} and RETRO~\cite{borgeaud2022improving}, Meta's MARGE~\cite{lewis2020pre}, Microsoft's GraphRAG~\cite{edge2024local}, and NVIDIA's InstructRETRO~\cite{wang2023instructretro}. As companies race to integrate RAG systems into their production pipelines~\cite{ruminer2024notebooklm,databricksrag,llamaindex}, optimizing their performance has become increasingly critical.

In contrast to conventional LLM-only serving systems, which center predominantly on optimizing the prefix (prompt decoding) and decoding (token generation) stages, RAG presents three challenges:
\textbf{(C1)} RAG systems are intrinsically heterogeneous, comprising a diverse array of system components, including vector search-based retrieval~\cite{borgeaud2022improving, shao2024scaling, lewis2020retrieval}, generative LLMs~\cite{llamaindex, team2024gemini, brown2020language}, and multiple optional models such as database encoders~\cite{reimers2019sentence, lee2024gecko}, query rewriters~\cite{chan2024rq, ma2023query}, and retrieval result rerankers~\cite{glass2022re2g, allahverdiyev2024chunkrag}.
These components often run on heterogeneous hardware platforms.
For example, retrievals are typically performed on CPU servers, whereas ML accelerators (e.g., TPUs or GPUs) are used for model serving.
This interplay of diverse components and hardware platforms amplifies the search space, far surpassing that of LLM-only systems;
\textbf{(C2)} Various RAG configurations defined by factors such as database size, retrieval frequency, model selection, and serving hardware, exhibit substantial performance variability.
This variability can veer the bottleneck between inference and retrieval or among different models within the serving pipeline; and
\textbf{(C3)} A natural consequence of the heterogeneity in components and the variability in performance is the emergence of a new challenge: \emph{how can we design efficient RAG serving systems?}
Addressing this challenge demands meticulously navigating key decisions in scheduling policies across diverse RAG configurations and hardware platforms.

To address these challenges in optimizing RAG serving performance, we ground our approach in three key design principles: 
(1) \textbf{Workload abstraction}: 
Tackling the heterogeneity of RAG systems necessitates an abstraction to encapsulate the diverse RAG workloads.
Without such abstraction, the inherent complexity of RAG configurations become intractable;
(2) \textbf{Critical system design decisions}: 
To unveil the critical system design decisions and illuminate the performance trade-offs inherent in RAG serving, a careful performance characterization of representative RAG workloads is warranted.
Without understanding how these different workloads behave, the optimization process risks becoming guesswork; and
(3) \textbf{Systematic optimization framework}: 
To navigate the large optimization space arising from the Cartesian product of RAG workload and system design dimensions, an optimization framework is essential to uncover and exploit efficiency opportunities in RAG serving systems.

To systematically describe RAG workloads, we introduce \ragdescemph (\S\ref{sec:rag-schema}), a RAG serving abstraction that encapsulates a set of essential performance-relevant workload attributes.
\ragdesc includes two key components: 
(a) specification of the RAG pipeline
\,---document encoder, query rewriter, result reranker, and generative LLM---\, and (b) model and retrieval configurations, including model size, database size, the number of query vectors per retrieval, and iterative retrieval frequency.
This abstraction simplifies the representation of complex RAG workloads while providing sufficient information for performance characterization and optimization.

Building on \ragdesc, we perform a detailed %
workload characterization (\S\ref{sec:case-studies}) to identify bottlenecks and key system design decisions.
We analyze four representative RAG paradigms, each with distinct RAG pipeline:
(a) RAG with hyperscale retrieval~\cite{borgeaud2022improving, shao2024scaling, wang2023instructretro};
(b) RAG for long-context sequence processing~\cite{lee2024can, li2024retrieval, yue2024inference};
(c) RAG with iterative retrieval~\cite{borgeaud2022improving, trivedi2022interleaving, jiang2023active}; and
(d) RAG with query rewriter and retrieval reranker models~\cite{chan2024rq, ma2023query, glass2022re2g, allahverdiyev2024chunkrag}.
Our analysis reveal \textit{significant performance variability both across and within paradigms}, with a subset of findings summarized as follows.
First, bottlenecks shift between retrieval and inference across RAG paradigms.
For instance, hyperscale retrieval can spend over 80\% in retrieval (\S\ref{subsec:case1}) while in long-context scenarios, retrieval accounts for less than 1\% of the total latency (\S\ref{subsec:case2}).
Second, even smaller models within the pipeline can significantly influence system performance.
For example, in long-context processing, a database encoder that is 100$\times$ smaller than the main generative LLM can become the bottleneck due to the large number of tokens it must process (\S\ref{subsec:case2}).
Third, iterative retrievals during decoding can stall the pipeline, as the decoding process waits for retrieval results (\S\ref{subsec:case3}).
The insights from these studies underscore not only the importance of making appropriate system design decisions but also the indispensable need for a tailored optimization framework for RAG serving systems, given their far less predictable performance landscape compared to LLM-only systems.
\begin{figure}[t]
  \centering
  \begin{subfigure}[b]{1.0\linewidth}
    \includegraphics[width=\linewidth]{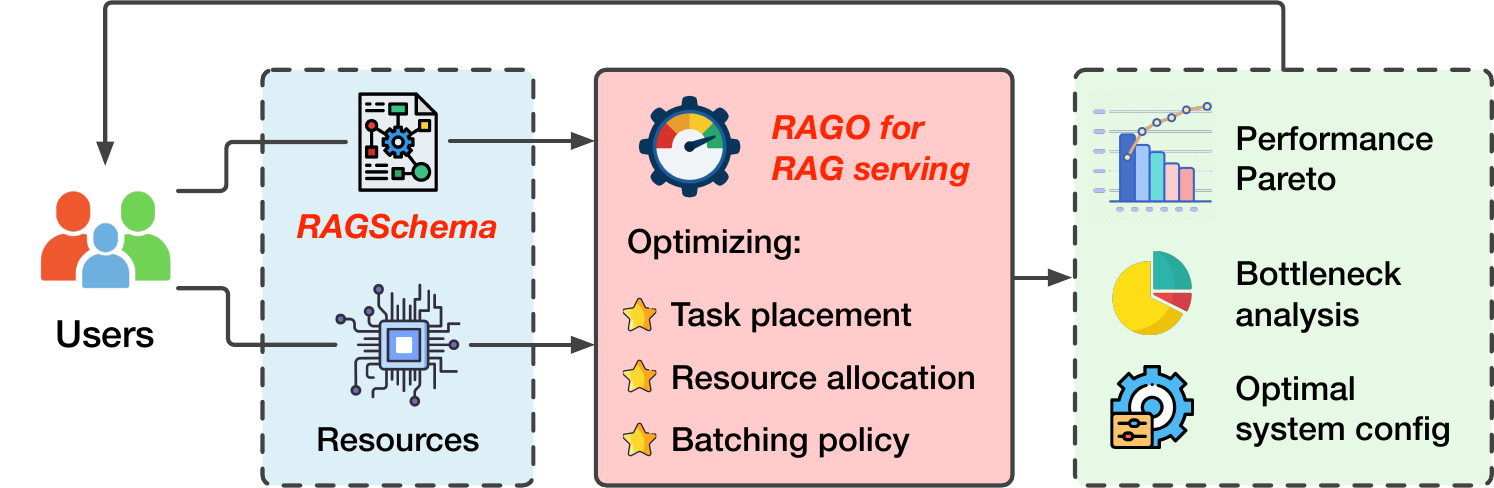}
  \end{subfigure}
  \vspace{-2em}
  \caption{\oursemph for systematic RAG serving optimization.}
  \vspace{-.5em}
  \label{fig:paper_overview}
\end{figure}

To this end, we introduce \oursemph~(\underline{R}etrieval-\underline{A}ugmented \underline{G}eneration \underline{O}ptimizer), a system performance optimization framework for efficient RAG serving (\Cref{fig:paper_overview}).
Given a RAG workload represented by \ragdesc and system resource constraints, this framework explores the scheduling policy space to determine optimal schedules aligned with user-defined performance objectives. 
Key scheduling decisions of \ours include deciding whether inference components are collocated or disaggregated across ML accelerators (\emph{task placement}), assigning the type and quantity of resources to each component (\emph{resource allocation}), and tuning batch sizes for retrieval and inference tasks to balance throughput and latency (\emph{batching policies}). 
\ours uses an analytical cost model, inspired by~\cite{parashar2019timeloop,zhang2022full,huang2024mind}, to identify the performance Pareto frontier and generate corresponding system schedules. 
This cost model is based on XPU, a generic systolic-array ML accelerator~\cite{jouppi2017datacenter,inferentia,zhang2022full}, and serve as the core engine of \ours for evaluating various RAG paradigms and configurations.
Below, we summarize the key contributions of our work:

\begin{itempacked}
\item We propose \ragdesc, a RAG workload abstraction that simplifies RAG workload representation and enables systematic performance characterization and optimization.
\item Using \ragdesc, we identify key system design decisions and performance trade-offs from characterizing four representative RAG paradigms and their instantiations. 
\item We develop \ours, a systematic optimization framework that optimizes scheduling policies for efficient RAG serving. Our results show that \ours delivers up to 2$\times$ improvement in QPS per chip and a 55\% reduction in time-to-first-token latency compared to RAG serving systems built on LLM-only systems.
\end{itempacked}

\section{Background}
\label{sec:background}

\niparagraph{Why RAG outshines LLM-only systems?}
LLM-only systems often struggle to achieve high factual accuracy or providing up-to-date information.
RAG addresses these limitations by combining the linguistic capabilities of LLMs with real-time knowledge retrieval~\cite{borgeaud2022improving, shao2024scaling, wang2023instructretro, lazaridou2021mind, lewis2020pre, xu2023nearest,
yu2024rankrag, yang2024crag, ram2023context}.
During offline pre-processing, external textual knowledge is encoded as vectors using an LLM and stored in a vector database.
At serving time, relevant knowledge is retrieved via vector search, assessing relevance by comparing the similarity between prompt's vector representation and those in the database.
The retrieved knowledge is then appended to the prompt, refining the quality of the LLM's responses.

By virtue of this combination, RAG systems offer several key advantages over LLM-only systems for knowledge-intensive tasks.
\textit{First}, RAG systems simplify knowledge updates by allowing external databases to be modified independently~\cite{borgeaud2022improving}, unlike LLMs, which require retraining or fine-tuning~\cite{borgeaud2022improving, lewis2020retrieval}.
\textit{Second}, RAG reduces ``hallucination''---a phenomenon where LLMs generate factually incorrect or entirely fictitious information.
For instance, a traditional LLM might confidently assert that ``Thomas Edison invented the internet'', despite this being obviously false.
RAG's reliance on external, up-to-date databases helps mitigate these errors by grounding the model's output in real, retrievable data~\cite{lewis2020retrieval, li2023dark}.
\textit{Finally}, RAG systems achieve comparable or better generation quality with models that are one to two orders of magnitude smaller than LLMs~\cite{lewis2020retrieval, izacard2020leveraging, komeili2021internet, guu2020retrieval, khandelwal2019generalization, khandelwal2020nearest, lewis2020pre}. 
While conventional LLMs require extensive parameters to encode a vast range of general knowledge~\cite{brown2020language, chowdhery2022palm, smith2022using, rae2021scaling}, RAG partially offloads this knowledge storage to an external database, retrieving only the most relevant content during inference.

\niparagraph{LLM-only serving systems.}
Serving LLM-only systems typically involves two distinct stages: prefix (prompt computation) and decode (token generation)~\cite{patel2023splitwise, zhong2024distserve}.
The prefix stage processes the input prompt to generate the first output token and populate the associated key-value (KV) cache~\cite{vaswani2017attention}, which holds the encoded representation of the input context.
The decode stage, on the other hand, generates subsequent tokens one at a time in an auto-regressive manner, relying on the KV cache from the prefix stage.

Modern LLM serving systems~\cite{patel2023splitwise, zhong2024distserve} often disaggregate these stages, running them on separate accelerators to accommodate their distinct workload characteristics, similar to disaggregated designs in recommender systems~\cite{ke2022disaggrec, jiang2021fleetrec}.
The prefix stage processes the entire input sequence at once, making it highly compute-intensive.
Even with small batches, the prefix stage benefits from accelerators with high computational throughput to handle the full sequence length efficiently~\cite{patel2023splitwise}.
In contrast, the decode stage is memory-bound, as each inference step requires accessing the KV cache of previous tokens, while the amount of computation is small~\cite{patel2023splitwise}.
In addition to workload differences, these two phases affect different performance metrics with different SLAs: time-to-first-token (TTFT) for the prefix phase and time-per-output-token (TPOT) for the decode phase.
Ultimately, optimizing the performance of LLM-only serving often depends on efficient resource allocation between the prefix and decode stages~\cite{patel2023splitwise}.

\niparagraph{Vector search for retrieval.}
Another core component in RAG systems is retrieval, which identifies information from external knowledge databases.
A common approach to performing this retrieval is \emph{vector search}, which has become the cornerstone of recent information retrieval systems~\cite{PQ, malkov2018efficient}.
Vector search enables the system to assess semantic relevance by encoding both documents and queries as high-dimensional vectors (\egc hundreds to thousands dimensions), where proximity in this vector space reflects semantic similarity.

In practice, vector search retrieves the $\mathcal{K}$ most similar vectors to a given $\mathcal{D}$-dimensional query vector $x$ from a database $\mathcal{Y}$ populated with many $\mathcal{D}$-dimensional vectors.
This similarity is typically computed using metrics such as L2 distance or cosine similarity~\cite{PQ, malkov2018efficient}.
Since exact $\mathcal{K}$ Nearest Neighbor (KNN) search is costly on large-scale datasets, real-world vector search systems adopt \emph{Approximate Nearest Neighbor} (ANN) search algorithms, which provide a scalable alternative to exact KNN by trading recall for much higher system performance.\footnote{We use ``vector search'' and ``ANN search'' interchangeably.}

The \emph{IVF-PQ} algorithm, which combines an inverted file (IVF) index with product quantization (PQ)~\cite{PQ}, is one of the most widely used approaches for large-scale vector search in RAG~\cite{borgeaud2022improving, izacard2020leveraging, khandelwal2019generalization}.
IVF-PQ is frequently preferred over other ANN algorithms, such as graph-based search algorithms~\cite{malkov2014approximate, malkov2018efficient, fu2017fast, zhao2023towards, zuo2023arkgraph, lu2021hvs, gao2023high}, due to its memory efficiency (e.g., one byte can represent 4$\sim$16 dimensions in PQ~\cite{johnson2019billion,PQ,jiang2023chameleon}) ---a crucial advantage when RAG systems operate on large databases, sometimes containing up to 64 billion vectors (92\,TB before quantization)~\cite{borgeaud2022improving, wang2023instructretro}.

Two popular open-source libraries for IVF-PQ are Faiss~\cite{faiss} and ScaNN~\cite{guo2020accelerating, scann_github}, exemplifying CPU-bound and memory-bound PQ variants, respectively.
Faiss employs a high-precision quantization variant, resulting in a CPU-bound search process~\cite{jiang2023chameleon, andre2016cache}.
In contrast, ScaNN adopts lower-precision quantization~\cite{sun2024soar}, achieving higher CPU throughput and shifting the workload toward being memory-bound.
\begin{table*}[t]
    \centering
    \setlength\dashlinedash{0.2pt}
    \setlength\dashlinegap{1.5pt}
    \setlength\arrayrulewidth{0.3pt}
    \begin{footnotesize}
    \setlength{\tabcolsep}{4pt} %
    \renewcommand{\arraystretch}{0.95} %
    \caption{\ragdesc component names, attributes, and corresponding example design parameters.}
  \vspace{-1em}
    \label{tbl:ragschema}
    \scalebox{1.0}{ %
        \begin{tabular}{p{3cm}!{\vrule width 0.7pt}p{11cm}!{\vrule width 0.7pt}p{2cm}!}
            \hline
            \textbf{\ragdesc Components} & \textbf{Attributes} & \textbf{Examples} \\ \hline
            Document Encoder & Model size (parameters) of the encoder used to convert database documents and queries into vector representations. & 120M \\ \hdashline
            Vector Dimensionality & The number of dimensions for each database vector. & 768-dim \\ \hdashline
            Database Vector Number & Number of the database vectors, depends on the corpus size and passage chunk lengths. & 1,000 \\ \hdashline
            Retrieval Frequency & Whether iterative retrievals are permitted during decoding and number of retrievals per sequence. & Four per sequence \\ \hdashline
            Queries Per Retrieval & Number of query vectors used per retrieval (one or multiple). & Two per retrieval \\ \hdashline
            Query Rewriter & Model size of the generative query rewriter, if applied. & 8B \\ \hdashline
            Query Reranker & Model size of the retrieval results reranker (usually an encoder-only model), if applied. & 120M \\ \hdashline
            Generative LLM & Represents the model size of the main generative LLM used for answer generation. & 70B \\ \hline
        \end{tabular}
    } %
    \end{footnotesize}
\end{table*}

\section{Structuring the Complex Terrain of RAG Serving}
\label{sec:rag-schema}
In this section, we first describe four representative RAG paradigms with increasingly diverse and complex RAG pipelines. 
We then describe \ragdesc~(\S\ref{sec:rag-schema:schema}), a structured abstraction to capture this workload diversity, serving as a foundation for serving performance characterization~(\S\ref{sec:case-studies}) and optimization~(\S\ref{sec:ragflow}).

\subsection{Representative RAG Paradigms}
\label{sec:rag-schema:cases}

We now show the workload diversity by describing the following representative RAG paradigms:

\niparagraph{Paradigm I: Hyperscale Retrieval.}
Retrieval over a large-scale corpus combined with smaller LLMs can serve as an alternative of larger LLMs without retrieval~\cite{borgeaud2022improving, shao2024scaling, wang2023instructretro}.
Prior work has shown that RAG systems can match or even surpass the quality of LLM-only systems when database sizes are sufficiently large~\cite{borgeaud2022improving, shao2024scaling}.
This is achieved while using sufficiently smaller models\,---approximately one-tenth the parameters of their LLM-only counterparts~\cite{borgeaud2022improving, wang2023instructretro}.
This quality parity is achieved because LLM-only models rely on their vast parameter sets to encode comprehensive knowledge during training~\cite{brown2020language, chowdhery2022palm, smith2022using, rae2021scaling}, whereas RAG systems dynamically integrate external knowledge at inference time, reducing the need for extensive parameterization within the model itself.

\niparagraph{Paradigm II: Long-Context Sequence Processing.}
Another common paradigm is to use RAGs to facilitate long-context processing~\cite{lee2024can, li2024retrieval, yue2024inference}.
For example, when answering questions based on a lengthy document (e.g., with more than 100K tokens) that a user has uploaded in real time --- similar to use cases in Gemini 1.5~\cite{team2024gemini}, NotebookLM~\cite{notebooklm}, and ChatGPT~\cite{chatgpt} --- a straightforward approach is to include the entire document in the prompt.
However, this approach is often prohibitively expensive due to the large number of tokens to process.
Instead, an efficient alternative is to treat the user-provided long document as a knowledge database, retrieving only the relevant information needed to answer the questions.
This method substantially reduces the prompt size by avoiding the need to load the full text into the model's context window.
Recent studies~\cite{lee2024can,yue2024inference} demonstrate that this retrieval-based approach achieves similar response quality to using the full document as a prompt, providing a practical balance between cost and quality in handling long contexts.
In contrast to the paradigm I, RAG for long-context processing introduces two key modifications.
First, this setup includes a database encoder, which is necessary for constructing the database when the long context is initially provided.
Second, the database is orders of magnitude smaller.
For example, given a context length of 100K tokens and a passage chunk size of 100 tokens, the database only consists of 1K vectors, compared to tens to hundreds of billions of vectors in paradigm I~\cite{borgeaud2022improving, shao2024scaling}.

\niparagraph{Paradigm III: Iterative Retrievals.}
While a single retrieval at the beginning may suffice in some scenarios, recent studies~\cite{borgeaud2022improving, trivedi2022interleaving, jiang2023active} indicate that iterative retrievals\,---periodically updating retrieved content during generation---\,can significantly enhance model quality.
Such update of the retrieved content is particularly valuable in scenarios requiring multi-hop reasoning, where each retrieval provides additional context to guide the subsequent token generation process~\cite{trivedi2022interleaving, yue2024inference}.
In this configuration, the decoder initiates retrievals at flexible intervals during generation.
Upon issuing a retrieval, the generation of this sequence temporarily pauses the token generation, to process newly retrieved content through the prefix phase.
Only after integrating this additional context does the decoder continue generating the rest of sequence.

\niparagraph{Paradigm IV: Query Rewriter and Reranker.}
Users often pose vague or complex queries, making it challenging to retrieve relevant information directly.
To address this, the retrieval process can be significantly improved by incorporating pre-processing and post-processing steps~\cite{chan2024rq, ma2023query, glass2022re2g, allahverdiyev2024chunkrag}.
For pre-processing, recent studies~\cite{chan2024rq, ma2023query} demonstrate that leveraging an LLM to rewrite the user's query can improve retrieval quality.
This LLM may either rephrase the query for clarity or decompose complex questions into multiple simpler queries that cover different aspects of the user's original intent~\cite{chan2024rq, ma2023query}.
Once the initial results are retrieved through vector search, a reranking model can be applied as a post-processing step~\cite{glass2022re2g, allahverdiyev2024chunkrag, rag_github}.
The reranker improves content retrieval quality by scoring each document's relevance beyond simple vector similarity and choosing documents that more closely align with the user's intended question.

\subsection{\ragdescemph for Workload Abstraction}
\label{sec:rag-schema:schema}
\begin{figure}[t]
  \centering
  \includegraphics[width=1.0\linewidth]{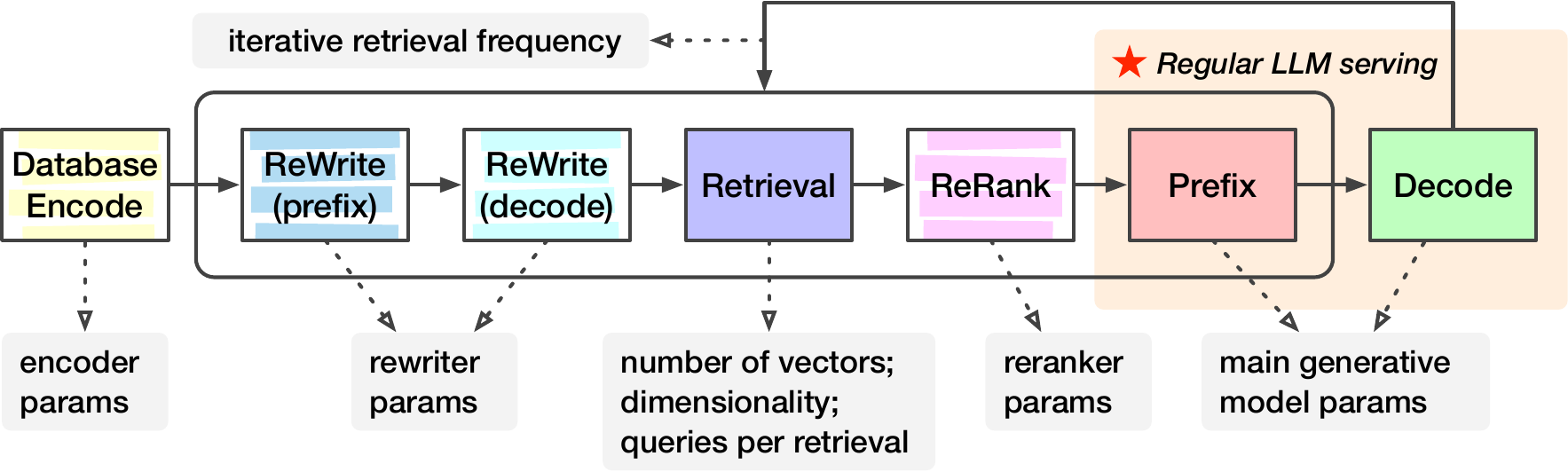}
  \vspace{-1.5em}
    \caption{Describing general RAG pipelines with \ragdesc.}
  \vspace{-.5em}
  \label{fig:rag-schema}
\end{figure}

Given these diverse paradigms, RAG workloads exhibit significant variability across algorithm configurations in the following ways.
First, retrieval configurations can vary dramatically. 
Database sizes may span several orders of magnitude (e.g., a million times)~\cite{borgeaud2022improving, shao2024scaling, lee2024can, li2024retrieval}; a retrieval may involve not a single query vector~\cite{lewis2020pre, guu2020realm} but multiple ones~\cite{wang2024richrag, besta2024multi, chan2024rq}; and some models support iterative retrievals during the generation process~\cite{borgeaud2022improving, trivedi2022interleaving, jiang2023active}.
Second, a RAG system may include several models in addition to the main generative LLM. 
These auxiliary models include a database encoder for processing real-time uploaded documents~\cite{reimers2019sentence, lee2024gecko}; a query rewriter model~\cite{chan2024rq, ma2023query} to rephrase user queries; and a result reranker model~\cite{glass2022re2g, allahverdiyev2024chunkrag, rag_github} to score retrieved information.

To navigate the complex RAG configuration space, we introduce \ragdescemph: \emph{a structured and modular abstraction that captures the key performance-relevant attributes of various RAG serving workloads}.
As visualized in \Cref{fig:rag-schema} and detailed in \Cref{tbl:ragschema}, \ragdesc defines both (1) the execution flow of the RAG pipeline and (2) the configuration of its components.
For the RAG pipeline definition, optional stages\footnote{A "stage" refers to the execution of a RAG pipeline component.}---such as the database encoder, query rewriter, reranker, and iterative retrieval---\,can be included or omitted.
For each included component, \ragdesc specifies relevant configurations, including model parameter counts, vector dimensionality, number of database vectors, queries per vector, and iterative retrieval frequency if applicable.
While \ragdesc abstracts RAG serving workloads, it is not an abstraction for quality, as different models and databases of the same size can lead to varying quality.

\subsection{Empirical RAG Performance Trade-off Analysis}
Even though precise bottlenecks and tradeoffs depend on exact RAGSchema, high-level performance bottlenecks in RAG systems are driven by RAG workload pipelines and the Amdahl's law.
In this section, we make general observations about RAG workloads, and quantify in ~\S\ref{sec:case-studies} using detailed performance models.

We represent inference throughput as a function of FLOPs, and retrieval throughput as a function of the bytes of database vectors accessed.
Note the precise throughput depends on CPU server efficiency, accelerator capability, scheduling policies, etc.

\niparagraph{Inference components.}  
For a model with size \( M \) and a sequence length \( L \), the FLOPs required for processing the entire sequence are approximately: \( \text{FLOPs}_{\text{inference}} \approx 2 \cdot M \cdot L \) for short sequences (e.g., \( L \leq 10^3 \)) where the quadratic complexity of the attention mechanism still has negligible impact.

\niparagraph{Retrieval component.}  
The retrieval workload can be approximately described by the number of bytes of database vectors processed per query. 
Unlike model inference, decoding quantized database vectors represents a fundamentally different workload where FLOPs is not an appropriate metric~\cite{jiang2023chameleon, andre2016cache}. 
Given a database with \( N_{\text{dbvec}} \) vectors, where each vector consists of \( B_{\text{vec}} \) bytes, and each query scans a subset of \( P_{\text{scan}} \) percent of database vectors, the total bytes to scan per query is approximately:  \(
\text{B}_{\text{retrieval}} \approx N_{\text{dbvec}} \cdot B_{\text{vec}} \cdot \frac{P_{\text{scan}}}{100}
\).
Here, \( P_\text{scan} \) is determined by evaluating a set of sample queries and analyzing the relationship between \( P_\text{scan} \) and retrieval quality measured by recall, as a common practice for retrieval configuration tuning~\cite{annbench}. The minimum value of \( P_\text{scan} \) that satisfies the required retrieval quality is then selected.

\niparagraph{End-to-end RAG performance.}  
While the latency of RAG serving is the sum of the latencies of each stage in the RAG pipeline, the throughput of the pipeline is determined by its slowest stage (excluding iterative retrieval paradigm for now, as it follows a different pattern discussed in~\S\ref{subsec:case3}).  
For a RAG pipeline with \( m \) stages, where each stage has a throughput denoted by \( \text{QPS}_i \) (\( i = 1, 2, \ldots, m \)), the end-to-end RAG serving throughput is:  
\(
\text{QPS}_{\text{RAG}} = \max(\text{QPS}_1, \text{QPS}_2, \ldots, \text{QPS}_m)
\).

From this high-level model, we can draw several key insights.
First, retrieval can become a bottleneck when its workload (\( N_{\text{dbvec}} \cdot B_{\text{vec}} \cdot \frac{P_{\text{scan}}}{100} \)) is high while the inference workload (\(  2 \cdot M \cdot L \)) is relatively low. 
Second, in paradigms with multiple inference components, any model can become critical depending on its size \( M \) and processed sequence length \( L \), which may vary based on the model's role.
Finally, the cumulative effect of multiple inference stages and retrievals can significantly impact overall serving performance.
We discuss detailed evaluation methodology and quantitative characterization in the subsequent sections.

\section{Methodology}
\label{sec:method}
This section outlines the methodology used to characterize RAG performance (\S\ref{sec:case-studies}) and evaluate \ours  across various configurations (\S\ref{sec:eval}).

\niparagraph{Models and database.}
We evaluate four LLMs---\bench{Llama-3 1B}, \bench{8B}, \bench{70B}, and \bench{405B}~\cite{dubey2024llama}---covering size scales comparable to those used in~\cite{shao2024scaling, borgeaud2022improving, wang2023instructretro}. 
We assume the models are quantized to 8-bit integer, thus the accelerator memory requirement directly corresponds to the model's parameter count (e.g., a 70B model requires 70 GB of memory).
As RAG quality continues to benefit from larger knowledge corpora~\cite{borgeaud2022improving, shao2024scaling}, we adopt a hyperscale database~\cite{borgeaud2022improving}.
This database contains 64 billion passages, each encoded as a 768-dimensional vector~\cite{borgeaud2022improving}, making it approximately 400$\times$ larger than the largest academic vector search datasets~\cite{SIFT, babenko2016efficient, simhadri2022results}.
We apply product quantization (PQ) as in~\cite{borgeaud2022improving} to compress each vector to 96 bytes (1 byte per 8 dimensions), resulting in a 5.6\,TiB quantized vector database. 
Following the index recommendations of the ScaNN library~\cite{scann_github}, we use a balanced fanout of 4K vectors per node across the three-level tree index~\cite{sun2023automating} ($(64 \times 10^9)^{1/3} = 4 \times 10^3$).
To balance retrieval quality and performance, each query is compared against 0.1$\%$ of the database vectors by default, as this setup has shown high recall (over 90\%) in billion-scale datasets~\cite{jiang2023chameleon}.

\niparagraph{LLM sequence lengths.}
In line with common RAG use cases such as question-answering~\cite{lewis2020pre, trivedi2022interleaving, chan2024rq}, we evaluate sequence lengths derived from QA datasets~\cite{bajaj2016ms, joshi2017triviaqa, rajpurkar2018know}, where the question lengths range from six to 42 tokens.
To simplify the search space, we use 32 tokens as the typical question length.
The input prompt length includes both the question and relevant retrieved content.
The typical nearest neighbor retrieved ranges from two to 10~\cite{asai2023self, jiang2023active, trivedi2022interleaving, borgeaud2022improving}, each with an average length of 100 tokens.
We pick five as a common value for our evaluations. 
Given this, we approximate the average length of input prompt (question + relevant retrieved contents) to 512 tokens.
For generation lengths (decode stage), we rely on data from long-form QA~\cite{fan2019eli5} and chatbot datasets~\cite{sharegpt, vllm}, selecting 256 tokens as a representative decode length. 

\begin{table}[!t]
\caption{Performance specifications of three versions of XPUs. We report performance on XPU-C (\textcolor{blue}{\(\star\)}) by default.}
  \vspace{-1em}
\label{tbl:xpu:gen}
\setlength\dashlinedash{0.2pt}
\setlength\dashlinegap{1.5pt}
\setlength\arrayrulewidth{0.3pt}
\begin{footnotesize}
\begin{center}
\setlength{\tabcolsep}{6pt} %
\renewcommand{\arraystretch}{1.2} %
\scalebox{1.0}{%
\begin{tabular}{@{\extracolsep{\fill}}l||c|c|c@{\extracolsep{\fill}}}
\Xhline{0.3ex}
& \textbf{XPU-A} & \textbf{XPU-B} & {\textcolor{blue}{\(\star\)}}\,\textbf{XPU-C}\\
\hline\hline
\textbf{TFLOPS}                & 197 & 275 & 459 \\
\textbf{HBM (GB)}              & 16 & 32 & 96 \\
\textbf{Mem. BW (GB/s)}        &  819     &  1200     & 2765 \\
\textbf{Inter-Chip Link BW (GB/s)} & 200   &  300     & 600 \\\hdashline
\textbf{Resembles} & TPU v5e~{\cite{tpuv5e}}& TPU v4~{\cite{tpuv4}} & TPU v5p~{\cite{tpuv5p}}\\\Xhline{0.3ex}
\end{tabular}
} %
\end{center}
\end{footnotesize}
\end{table}

\niparagraph{System setup.}
Our evaluation assumes a data center model serving environment with abundant resources to support various system configurations.
Across the RAG serving stages (\egc prefix, decode), we allocate a total of 16 to 32 servers hosting 64 to 128 XPUs (4 XPUs per server), as a minimum of 16 servers is required to ensure sufficient host memory capacity for the dataset (5.6\,TiB after quantization).
An XPU refers to a generic systolic-array-based ML accelerator~\cite{jouppi2017datacenter, inferentia}.
The number of XPUs allocated to each model component is configured in powers-of-two scaling factors (e.g., 1, 2, 4, etc.).
Each XPU, inspired by the setup of TPU v5p accelerators~\cite{tpuv5p}, is equipped with 96\,GB of high-bandwidth memory (2.7\,TB/s) and 459\,TFLOPS of compute capacity.
The XPUs are interconnected via a high-bandwidth 3D torus topology, offering 600\,GB/s of inter-chip bandwidth (six 100\,GB/s links per chip).
We also evaluate two other versions of XPUs, as shown in Table~\ref{tbl:xpu:gen}, for ablation studies.
The host CPUs are modeled after AMD EPYC Milan processors, featuring 96 cores, 384\,GB of memory, and 460\,GB/s of memory bandwidth.
We assume that XPU host servers support distributed retrieval across large databases.

\niparagraph{Simulation setup.}
RAG performance is reported by assembling the costs of all model inference and retrieval stages, based on a search across various system configurations (details described in~\S\ref{sec:ragflow}).
We now describe the production-grade simulators used to measure inference and retrieval performance.
With the following simulation mechanism, we can model RAG serving performance given XPUs and CPU servers with varying configurations, as well as different distributed serving configurations in data centers.

\begin{figure}[t]
  \centering
  \begin{subfigure}[b]{0.8\linewidth}
    \includegraphics[width=\linewidth]
    {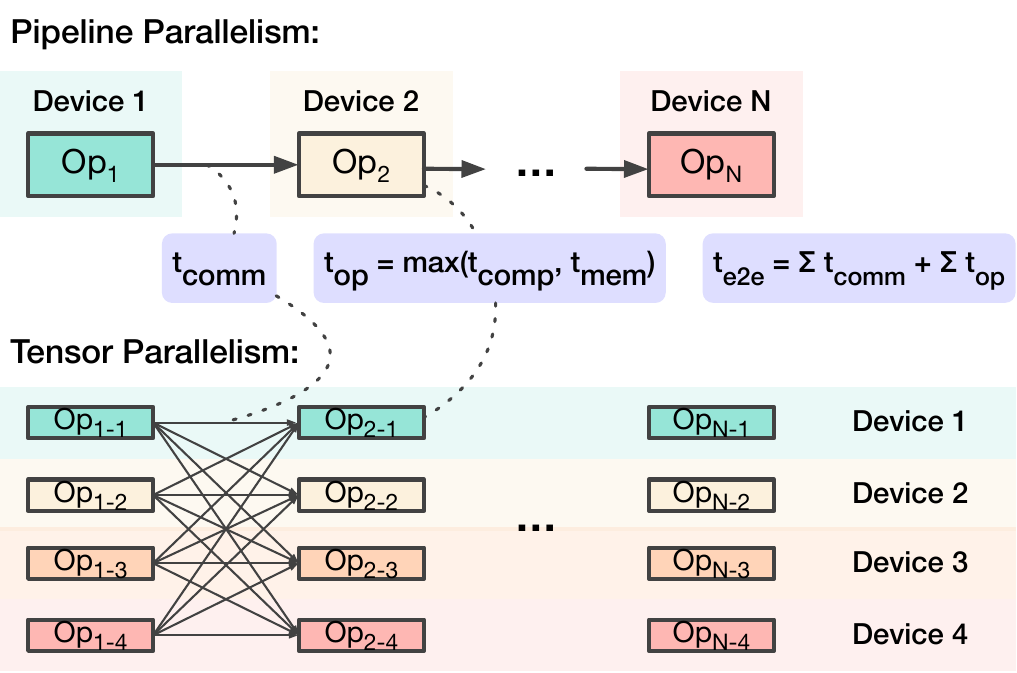}
  \end{subfigure}
    \vspace{-0.5em}
  \caption{Parallelisms and cost model in model inference.}
  \vspace{-.5em}
    \label{fig:roofline}
\end{figure}

\emph{(a) Inference performance modeling.}
We adopt an in-house calibrated XPU simulator for inference simulation. 
The simulator is well-correlated with the production-grade XPU accelerators across a set of real-world ML models.
For multi-XPU inference, the simulator evaluates a range of model sharding strategies, where each accelerator unit is assigned a subset of inference operators.
The simulator supports pipeline parallelism~\cite{huang2019gpipe, narayanan2019pipedream}, tensor parallelism~\cite{shoeybi2019megatron, rajbhandari2020zero}, and hybrid approaches. 
As shown in \Cref{fig:roofline}, the simulator abstracts inference as a sequence of operators, following a methodology similar to other established ML simulators~\cite{zhang2022full,parashar2019timeloop}.
That is, the total latency is computed as the sum of each operator’s execution time and the associated communication costs between operators. 
The execution time of each operator is calculated using a roofline model, where latency is determined by the maximum of memory access latency and compute latency:
\( T_{op_i} = \max \left( \frac{F_i}{P_{\text{comp}}(F_i)}, \frac{D_i}{B_{\text{mem}}(D_i)} \right) \),  
where \( F_i \) is the number of floating-point operations required by operator \( op_i \), \( P_{\text{comp}} \) represents the compute performance, \( D_i \) denotes the total data size processed by \( op_i \), and \( B_{\text{mem}} \) is the available memory bandwidth.  
Additionally, the communication latency between two operators depends on the volume of data transferred and the network bandwidth:  
\( T_{comm}(op_i, op_j) = \frac{S_{i,j}}{B_{\text{net}}} \), where \( S_{i,j} \) represents the size of data transferred between \( op_i \) and \( op_j \), and \( B_{\text{net}} \) is the available network bandwidth (bytes/second).

\emph{(b) Retrieval performance modeling.}
Our retrieval simulation is based on ScaNN~\cite{guo2020accelerating, scann_github}, a product quantization library~(\S\ref{sec:background}) that demonstrates state-of-the-art performance across dozens of algorithms in the ANN benchmark~\cite{annbench}.
We implement the ScaNN performance model described in~\cite{sun2023automating}, which models the search process as a sequence of vector scan operations at each level of a multi-level tree~\cite{sun2023automating, scann_github}. The total retrieval latency is calculated as the sum of the latencies for these scan operations.
ScaNN dedicates one thread per query and parallelizes batches of queries across multiple threads.
The cost of each operator is calculated by a roofline model that factors in batch sizes, the number of CPU cores, per-core CPU processing throughput, and memory bandwidth.
The execution time of each scan operator is determined by the maximum of compute time and memory access time:  
\(
T_{op_i} = \max \left( \frac{D_i}{P_{\text{comp}}(Q)}, \frac{D_i}{B_{\text{mem}}(D_i)} \right)
\), 
where \( D_i \) represents the total number of bytes processed by operator \( op_i \), \( Q \) denotes the query batch size, \( P_{\text{comp}} \) is the CPU throughput for handling the scan operation, and \( B_{\text{mem}} \) represents the memory bandwidth.

For large databases requiring distributed search across multiple servers, we assume each server holds a shard of the dataset with independent indexes.
Queries are routed to all servers, and results are aggregated. 
The workload is balanced across servers, with negligible overhead for broadcast and gather operations.

To populate simulator parameters, we benchmark the maximum achievable per-core throughput and memory bandwidth by running open-source ScaNN~\cite{scann_github} on smaller datasets configured with the same tree node sizes (4K vectors per node) as the 64-billion vector database.
On AMD EPYC 7R13 CPUs with 24 cores, ScaNN achieved a PQ code scanning throughput of 18\,GB/s per CPU core, with approximately 80\% memory bandwidth utilization.
We then calibrate the retrieval performance model using internal production datasets comparable in scale to the 64-billion vector dataset used in our study~(\S\ref{sec:method}).

\emph{(c) Communication between retrieval and inference.}  
Retrieved documents are transferred from CPUs to XPUs, with each retrieval’s data size modeled as \( N_{\text{token}} \times B_{\text{token}} \), where \( N_{\text{token}} \) represents the number of tokens and \( B_{\text{token}} \) denotes the bytes per token.  
In practice, this communication overhead is negligible.  
For instance, retrieving five documents, each containing 100 tokens at 2 bytes per token, results in only 1 KB of data transmission per query.  
Given a typical PCIe bandwidth of tens of GB/s, a system can support tens of millions of requests per second --- several orders of magnitude higher than the typical LLM inference throughput of up to tens of requests per second per XPU, as shown in \Cref{fig:case1:rag_vs_llm}.  

\begin{table}[t]
    \centering
    \setlength\dashlinedash{0.2pt}
    \setlength\dashlinegap{1.5pt}
    \setlength\arrayrulewidth{0.3pt}
    \begin{footnotesize}
    \setlength{\tabcolsep}{4pt} %
    \renewcommand{\arraystretch}{0.95} %
    \caption{\ragdesc of the workloads used in case studies.}
    \vspace{-1em}
    \label{tbl:ragschema-example}
    \scalebox{0.92}{ %
        \begin{tabular}{p{2.8cm}!{\vrule width 0.7pt}p{1.5cm}!{\vrule width 0.7pt}p{1.5cm}!{\vrule width 0.7pt}p{1.cm}!{\vrule width 0.7pt}p{1.cm}}
            \hline
            \textbf{\ragdesc Components} & \textbf{Case 1} & \textbf{Case 2} & \textbf{Case 3} & \textbf{Case 4} \\ \hline
            Document Encoder & N/A & 120M (768-d)  & N/A & N/A \\ \hdashline
            Database Vector Number & 64B & 1/10/100K & 64B & 64B  \\ \hdashline
            Retrieval Frequency & 1 & 1 & 2/4/8 & 1 \\ \hdashline
            Queries Per Retrieval & 1/2/4/8 & 1 & 1 & 1 \\ \hdashline
            Query Rewriter & N/A & N/A & N/A & 8B \\ \hdashline
            Query Reranker & N/A & N/A & N/A & 120M  \\ \hdashline
            Generative LLM & 1/8/70/405B & 8/70B & 8/70B & 8/70B \\ \hline
        \end{tabular}
    } %
    \end{footnotesize}
  \vspace{-1em}
\end{table}

\niparagraph{Performance metrics.}
We report common metrics used in the evaluation of LLM systems~\cite{patel2023splitwise,zhong2024distserve}:
\begin{itempacked}
\item {[\textbf{TTFT}]\,\textbf{Time-to-First-Token}} $\mapsto$ average latency from request reception to the generation of the first output token.
\item {[\textbf{TPOT}]\,\textbf{Time-Per-Output-Token}} $\mapsto$ average latency between generating each output token in a sequence. 
\item {[\textbf{QPS}]\,\textbf{Queries-Per-Second}} $\mapsto$ maximum throughput, or the number of requests the system can process per second. \footnote{Note that the term ``queries'' here does not refer to retrieval queries, and we use the QPS metric exclusively for end-to-end RAG serving performance in this paper.}
\item {[\textbf{QPS/Chip}]\,\textbf{Queries-Per-Second/Chip}} $\mapsto$ QPS normalized by chip count, reflecting system cost efficiency. 
\end{itempacked}

Since continuous batching~\cite{yu2022orca, vllm} is enabled in the decode stage, we report the worst-case TPOT latency.
This is because sequences in the batch can be at different stages of generation --- some generating the first token and others generating the last ones --- and performance is determined by the latter.
In contrast, prefix operates deterministically, allowing us to report the precise TTFT latency.

\section{RAG Serving Performance Characterization}
\label{sec:case-studies}

In this section, we characterize workloads using four case studies, each representing a \ragdesc instantiation of a distinct RAG paradigm described in~\S\ref{sec:rag-schema:cases}.
These case studies highlight the performance variability across RAG workloads, quantify the impact of paradigm choices on system performance, and motivate the need for our RAG optimization framework~(\S\ref{sec:ragflow}).
While arbitrary RAG configurations can be constructed beyond these studies, they can be seen as interpolations of the provided cases. 
For example, a RAG system with a small private database can be seen as a hybrid of Case I and II, where the large-scale retrieval in Case I is replaced with the small-scale retrieval of Case II. Similarly, long-context processing with iterative retrieval can be viewed as a combination of Case II and III.

\niparagraph{Characterization methodology.}
We evaluate performance using the methodology outlined in~\S\ref{sec:method}.
Unless otherwise specified, the end-to-end performance plots depict the performance Pareto across all system scheduling options.
The time breakdown plots are normalized by the resource usage of each component, reflecting time $\times$ resource consumption. 
These plots assume (a) four XPUs per host server and (b) each component operating at its maximum QPS/Chip.
Thus, if a plot shows that retrieval time exceeds the combined time of all inference components, the host servers for retrievals are the bottleneck, leaving XPUs occasionally idle; conversely, if inference dominates, retrieval resources may be underutilized.
The evaluated workloads are summarized in~\Cref{tbl:ragschema-example}. 
While all configurations in the table are analyzed, the plots highlight a representative subset to avoid redundancy when similar trends are observed across multiple configurations (e.g., model sizes).

\subsection{{\textup{\textbf{Case I:}}} Hyperscale Retrieval}
\label{subsec:case1}
As introduced in \S\ref{sec:method}, we adopt configurations similar to those in RETRO~\cite{borgeaud2022improving}, replicating its database setup and similar sized LLMs along with more recent, larger models.
The RAG system performs only \emph{one} retrieval at the beginning, which may involve one or multiple query vectors, as suggested by recent studies~\cite{wang2024richrag, besta2024multi, chan2024rq}.

\begin{figure}[t]
  \centering
  \begin{subfigure}[b]{0.8\linewidth}
    \includegraphics[width=\linewidth]
    {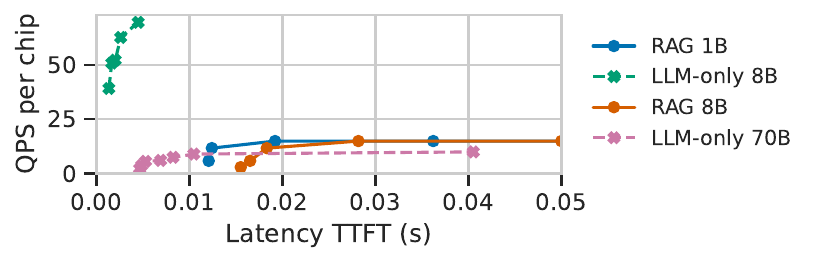}
  \end{subfigure}
  \vspace{-1em}
  \caption{Larger LLM versus RAG with smaller models.}
  \vspace{-1em}
    \label{fig:case1:rag_vs_llm}
\end{figure}

\niparagraph{Takeaways:} 
Hyperscale retrieval can pose a significant bottleneck in RAG pipelines. 
This bottleneck becomes increasingly dominant with (1) smaller LLM, (2) multi-query retrievals, (3) better inference accelerators, (4) shorter prefix and decode sequence lengths, and (5) higher retrieval quality.

\niparagraph{System performance comparison (RAG vs. LLM-only). }
\Cref{fig:case1:rag_vs_llm} compares RAG and LLM-only systems across different model sizes, with TTFT latency on the x-axis and QPS/Chip on the y-axis.
As shown in the RETRO paper~\cite{borgeaud2022improving}, RAG can achieve similar or superior generation quality to LLM-only systems with an order of magnitude fewer parameters.
Here, we extend this comparison to system performance.
Our results indicate that RAG 8B outperforms LLM-only 70B in QPS/Chip by a factor of 1.5$\times$.
Although the model size is reduced by approximately 10$\times$, the benefits of using smaller models in RAG are moderated by the retrieval overhead and the need for longer prompts to integrate retrieved information (512 tokens in RAG versus 32-token questions in LLM-only systems), resulting in only a 3.2$\times$ reduction in inference FLOPs.
Consequently, the QPS/Chip gain is not directly proportional to the reduction in parameter size.
Interestingly, the results suggest that RAG model sizes can be increased up to a certain limit without compromising QPS/Chip, as retrieval performance is the limiting factor.
For example, RAG 1B and RAG 8B exhibit similar QPS, highlighting the importance of system performance analysis in determining how much larger RAG models can scale.
While RAG models offer significant advantages at certain scales, their benefits may diminish at lower parameter counts as retrieval latency becomes a bottleneck. 
For example, despite RAG 1B having only one-eighth the parameters of LLM-only 8B, its QPS/Chip does not scale proportionally, because the retrieval overhead in RAG outweigh the benefits of reduced model size.

\begin{figure}[t]
  \centering
  \begin{subfigure}[b]{0.49\linewidth}
    \includegraphics[height=10em]
    {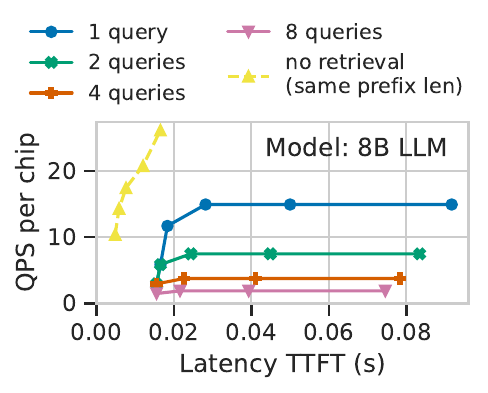}
    \caption{QPS/Chip 8B}
    \label{fig:case1:latency:8b}
  \end{subfigure}
  \begin{subfigure}[b]{0.49\linewidth}
    \includegraphics[height=10em]
    {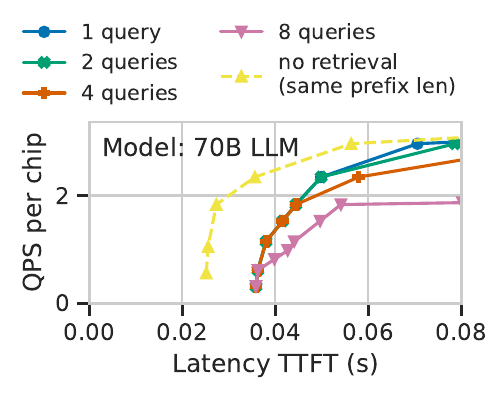}
    \caption{QPS/Chip 70B}
    \label{fig:case1:latency:70b}
  \end{subfigure}
  \begin{subfigure}[b]{0.49\linewidth}
    \includegraphics[width=\linewidth]
    {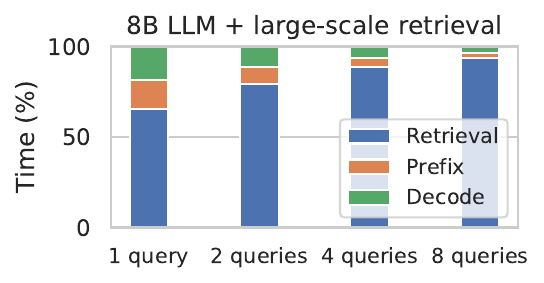}
    \caption{Breakdown 8B}
    \label{fig:case1:break:8b}
  \end{subfigure}
  \begin{subfigure}[b]{0.49\linewidth}
    \includegraphics[width=\linewidth]
    {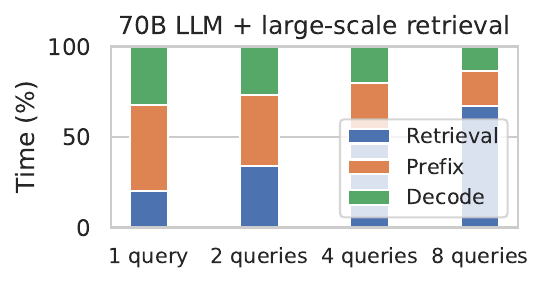}
    \caption{Breakdown 70B}
    \label{fig:case1:break:70b}
  \end{subfigure}

  \vspace{-1em}
  \caption{RAG performance given various model size and query numbers for hyperscale retrieval.}
  \vspace{-1em}
  \label{fig:case_1_latency_throughput}
\end{figure}

\niparagraph{Sensitivity to model size.}
Figure~\ref{fig:case1:latency:8b} and Figure~\ref{fig:case1:latency:70b} present the QPS/Chip for the \bench{8B} (left) and \bench{70B} (right) models, alongside time breakdowns for retrieval, prefix, and decode stages in Figure~\ref{fig:case1:break:8b} and Figure~\ref{fig:case1:break:70b}.
The yellow line represents a ``no retrieval'' configuration, where retrieval is omitted while the prefix remains the same length.
For the 8B model, retrieval is the primary bottleneck; as query counts double, QPS nearly halves due to increased retrieval demands.
Conversely, for the 70B model, inference initially limits performance until four queries per retrieval.
At higher query vector counts per retrieval (\egc 8 queries), the bottleneck shifts, and retrieval starts to dominate, as seen in the time breakdown in Figure~\ref{fig:case1:break:70b}.

\begin{figure}[t]
  \centering
  \begin{subfigure}[b]{0.3\linewidth}
    \includegraphics[height=6.2em]
    {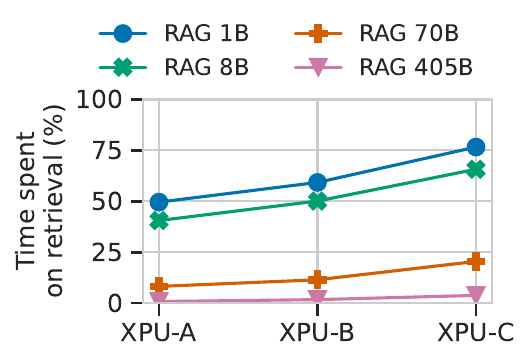}
    \caption{XPU Gen}
    \label{fig:case1:xpu}
  \end{subfigure}
  \begin{subfigure}[b]{0.3\linewidth}
    \includegraphics[height=6.2em]
    {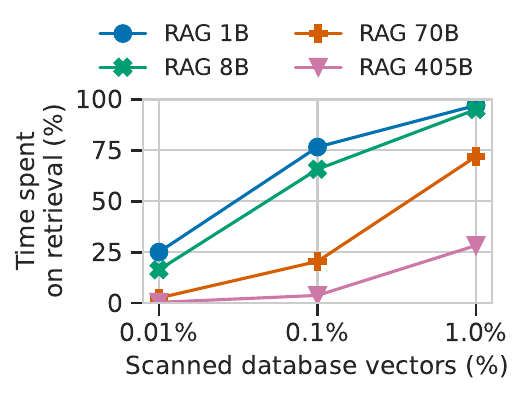}
    \caption{Retrieval Config}
    \label{fig:case1:db_scan}
  \end{subfigure}
  \begin{subfigure}[b]{0.38\linewidth}
    \includegraphics[height=6.2em]
    {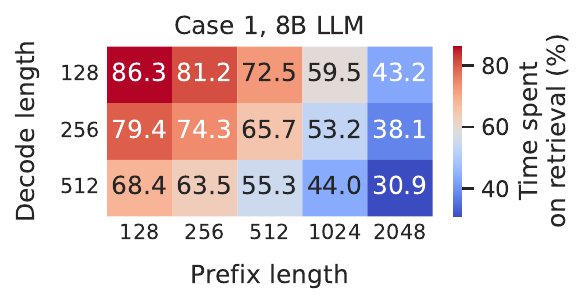}
    \caption{Seq. Length}
    \label{fig:case1:seqlen}
  \end{subfigure}
  \vspace{-1em}
  \caption{The percentage of retrieval time across hardware, retrieval configurations, and sequence lengths in Case I. }
  \vspace{-.5em}
  \label{fig:case_1_sweep_hardware_models}
\end{figure}

\niparagraph{Sensitivity to XPU versions.}
\Cref{fig:case1:xpu} shows the impact of XPU capability on the percentage of time spent on retrieval for LLMs ranging from 1B to 405B parameters.
As the XPU capabilities advance (from version A to C), the proportion of time spent on retrieval increases by up to 25\%.
While for larger models (e.g. 405B), LLM remains the dominant bottleneck in RAG serving, retrieval is dominant factor for RAG with small models (50\% - 75\% across XPUs versions).
Overall, with more advanced ML accelerators, system efficiency increasingly depends on optimizing retrieval processes.

\niparagraph{Sensitivity to sequence lengths.}
\Cref{fig:case1:seqlen} illustrates the sensitivity of retrieval overhead to changes in decode length and prefix length given for \bench{8B} model.
The retrieval overhead varies significantly with both decode and prefix lengths --- retrieval bottlenecks diminish as sequence lengths increase, shifting retrieval from a primary bottleneck to a secondary factor.
For example, 86.3\% of the time is spent on retrieval at  shorter sequence lengths (\egc 128 or 256), while the retrieval overhead drops to just 30.9\% with longer prefix and decode lengths (2048 and 512).
Adjusting the prefix and decode lengths results in unequal changes in the percentage of retrieval time.
For example, in a setting of 128 tokens for both prefix and decode, increasing the prefix length to 256 tokens reduces retrieval time from 86.3\% to 81.2\%, while increasing the decode length to 256 tokens lowers it to 79.4\%.
This difference occurs because prefix inference is inherently faster than decoding the same number of tokens due to the autoregressive nature of decoding.

\niparagraph{Sensitivity to retrieval configurations.}
Retrieval performance in RAG workflow is highly sensitive to the percentage of database vectors scanned per search.
Regardless of the ANN algorithm used, ANN search does not conform to a fixed workload --- there is an fundamental trade-off between retrieval performance and quality: scanning more vectors improves quality but reduces performance~\cite{PQ, malkov2018efficient}.
This trade-off is further influenced by data distribution; for instance, with the same algorithm, hardware, and QPS, one dataset may achieve over 90\% recall, while another may fall below 50\%~\cite{simhadri2022results}.
While prior evidence suggests that higher recall can enhance generation quality~\cite{jiang2024piperag, 2024ragannquality}, there has been no consensus on the optimal recall threshold.
\Cref{fig:case1:db_scan} illustrates the impact of varying the percentage of scanned database vectors, ranging from 0.01\% to 1\% (with 0.1\% as the default), on the proportion of time spent on retrieval across different model sizes.
For all models, increasing the scanned database vectors significantly amplifies the proportion of time spent on retrieval, highlighting the substantial variability in retrieval performance across RAG configurations.

\subsection{{\textup{\textbf{Case II:}}} Long-Context Sequence Processing}
\label{subsec:case2}
\begin{figure}[t]
  \centering
  \begin{subfigure}[b]{0.45\linewidth}
    \includegraphics[height=7em]
    {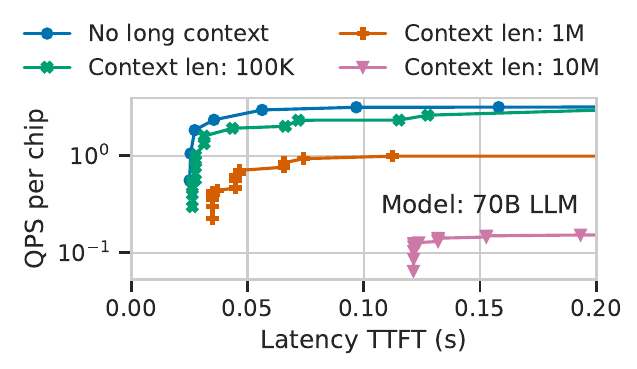}
    \caption{QPS/Chip 70B}
    \label{fig:case2:latency:70b}
  \end{subfigure}
  \begin{subfigure}[b]{0.54\linewidth}
    \includegraphics[height=7em]
    {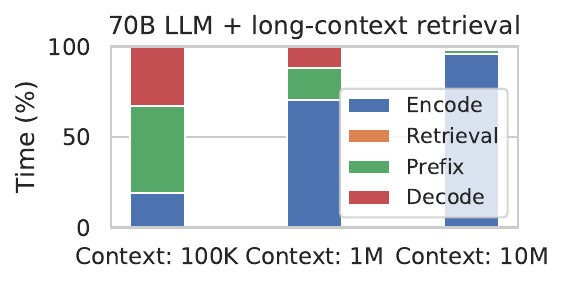}
    \caption{Breakdown 70B}
    \label{fig:case2:break:70b}
  \end{subfigure}
  \vspace{-1.5em}
  \caption{RAG performance for long-context processing.}
  \vspace{-.5em}
  \label{fig:case_2_latency_throughput}
\end{figure}

As shown in \Cref{tbl:ragschema-example}, we evaluate context lengths ranging from 100K to 10M tokens, resulting in a database of 1K to 100K vectors, with each chunk sized at 128 tokens and small overlaps between chunks.
We use a sentence transformer model with 120\,M parameters~\cite{reimers2019sentence} to encode the passages, generating 768-dimensional embeddings, as relatively compact models are sufficient to achieve high retrieval quality~\cite{reimers2019sentence}.
Instead of ANN search, we use brute-force kNN search due to the high indexing costs associated with newly generated embeddings.

\niparagraph{Takeaways:}
In contrast to the Case I, retrieval performance plays a minimal role here.
Instead, the database vector encoding process emerges as the bottleneck, even with a small encoder model, due to the significantly longer context the encoder must process compared to the generative LLM.

\niparagraph{Sensitivity to context length.}
Figure~\ref{fig:case_2_latency_throughput} presents performance trends when the input context length scales from 100K to 10M tokens for the 70B model.
"No long context" line represents the standard prompt length of a 512-token prefix.
As the context length increases, RAG performance gradually degrades due to the increasing cost of context encoding, even though retrieval enables prompt truncation for the generative LLM.
This happens due to database encoding becoming the bottleneck (Figure~\ref{fig:case_2_latency_throughput}), especially at longer context lengths ($>$1M).
Notably, encoding time scales with context length, despite the relatively small encoder applied (120M parameters), due to the sheer volume of data processed.
Therefore, caching the generated embedding for potential reuse can significantly reduce computation with minimal cost.
For instance, caching 10K 768-d database vectors in FP16 format (for 1M tokens) requires only 15 MB of CPU memory or storage.
The retrieval time is minimal (0.01\% to 0.4\% of the end-to-end latency) even when using brute-force search due to small database (1K-100K vectors vs 64B in other cases).

\niparagraph{RAG vs. long-context LLMs.}
Despite the high encoding cost, RAG is significantly more efficient than processing the entire long context as a prompt (long-context LLM).
For instance, with a 1M-token context and a 70B LLM, RAG reduces the required prefix length to 512 tokens, achieving a speedup of 2852.6$\times$ in TTFT and 6633.9$\times$ in QPS/Chip.
This is even considering an efficient long-context LLM applying global attention to all tokens in only one out of every four layers, while the rest layers only apply local attention to the last 128 tokens.
This cost efficiency arises for two primary reasons: ($\mathrm{I}$) In long-context RAG, the database encoder, typically a small model (\egc 120M parameters), performs the encoding.
This is much less computationally intensive compared to the LLM-only system with billions of parameters, which would require significantly more FLOPs if fed the entire context.
($\mathrm{II}$) 
Long-context LLMs require key-value caches for every token, consuming substantial XPU memory (i.e. cost).
In contrast, RAG significantly reduces prompt lengths, saving XPU memory.
This distinction enables RAG to handle larger batch sizes during generation, increasing QPS/Chip.

\subsection{{\textup{\textbf{Case III:}}} Iterative Retrievals + Prefix}
\label{subsec:case3}
\begin{figure}[t]
  \centering
  \begin{subfigure}[b]{0.48\linewidth}
    \includegraphics[width=\linewidth]
    {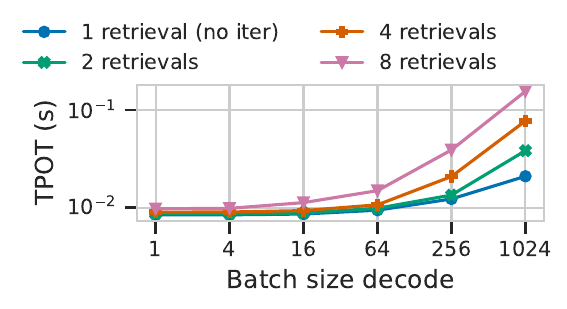}
    \caption{Retrieval frequency}
    \label{fig:case_3_latency_throughput_retrievals}
  \end{subfigure}
  \begin{subfigure}[b]{0.48\linewidth}
    \includegraphics[width=\linewidth]
    {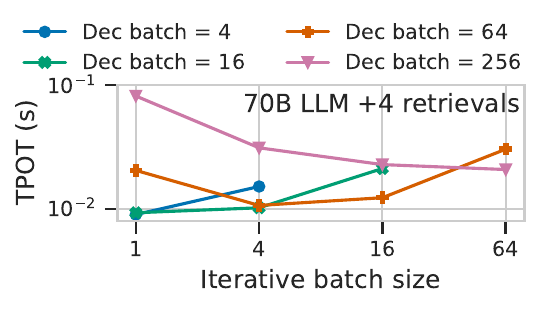}
    \caption{Prefix-retrieval batch size}
    \label{fig:case_3_latency_prefix_decode_batch}
  \end{subfigure}
    \vspace{-1em}
  \caption{RAG performance with iterative retrievals.}
    \vspace{-0.5em}
  \label{fig:case_3_latency_throughput}
\end{figure}

Building on Case I with hyperscale retrieval, this study adopts an iterative retrieval setup, allowing for 2, 4, or 8 retrievals per sequence generation process.
Each retrieval is triggered at random intervals during the 256-token decoding process, with retrievals uniformly distributed across token positions.

\niparagraph{Takeaways:}
Batch sizes for iterative retrievals must be carefully selected, as they significantly impact TPOT latency.
Larger batch sizes improve retrieval and prefix throughput but may stall decoding. %

\niparagraph{Sensitivity to retrieval frequency.}
\Cref{fig:case_3_latency_throughput_retrievals} examines the impact of different retrieval frequencies (1-8 per sequence) on TPOT latency as the decode batch size increases from 1 to 1024, as QPS/Chip shows similar trends as multi-query retrieval in Case I.
The results indicate that TPOT latency increases with both retrieval frequency and the decode batch size.
At smaller decode batch sizes (one, four, and 16), the TPOT latency differences between retrieval frequencies are relatively minor.
This is because, at these lower batch sizes, the decode step remains the dominant factor in TPOT latency, contributing approximately 60\%-80\% of the latency, while the effect of additional retrievals remains limited.
At higher batch sizes, however, the decode process achieves higher QPS/Chip, reducing its share of the overall TPOT latency. 
This shift in bottleneck exposes the impact of retrieval frequency, as retrievals become the primary contributor to latency.
Consequently, at larger batch sizes, the latency gap across different retrieval frequencies widens, making the increased time required for multiple retrievals more pronounced.

\niparagraph{Sensitivity to iterative retrieval batch size.}
In \Cref{fig:case_3_latency_prefix_decode_batch}, we observe the nuanced interplay between decode batch size, iterative retrieval-prefix batch size and TPOT latency for a 70B model processing four retrievals per sequence.
At smaller decoding batch sizes (4 and 16), increasing the iterative retrieval batch size results in a noticeable increase in latency.
This is due to the inherent challenge in finding enough retrieval requests to batch within an active set of decoding sequences over a given time interval, introducing stalls.
For decode batch sizes of 256, the relationship reverses.
As the iterative retrieval batch size increases, latency decreases.
Here, the abundance of active decode sequences allow the system to batch retrieval requests more rapidly, enabling improved performance.
The decode batch size of 64 presents a particularly intriguing case: it reaches its lowest TPOT at retrieval batch size of four.
This optimal point represents a balance where idle time is minimized and the batching of retrieval requests is most efficient.
However, beyond this threshold, latency begins to climb again as it becomes progressively harder to amass a sufficient number of retrieval requests for efficient batching.
This behavior illustrates the delicate balance in RAG system performance when trying to balance retrieval performance and decoding efficiency.

\begin{figure}[t]
  \centering
  \begin{subfigure}[b]{0.48\linewidth}
    \includegraphics[height=10em]
    {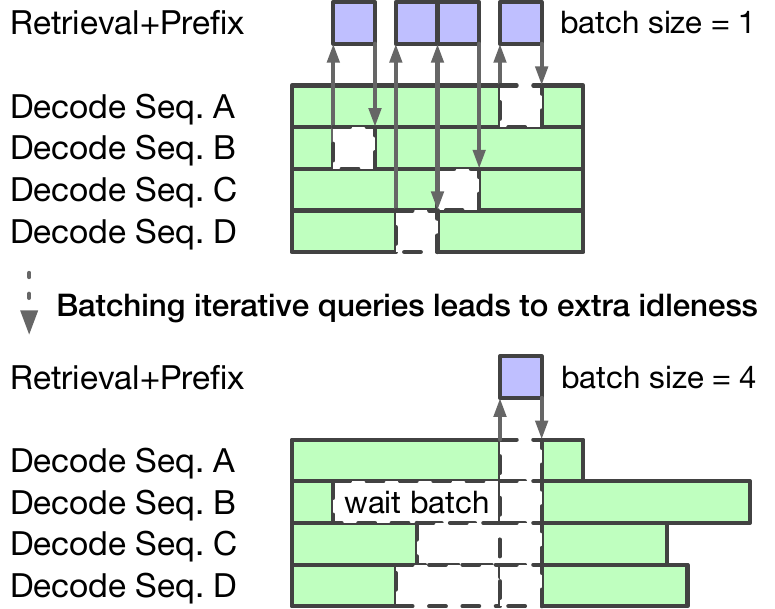}
    \caption{Wait for query batching}
    \label{fig:case_3_batching_idleness}
  \end{subfigure}
  \begin{subfigure}[b]{0.48\linewidth}
    \includegraphics[height=10em]
    {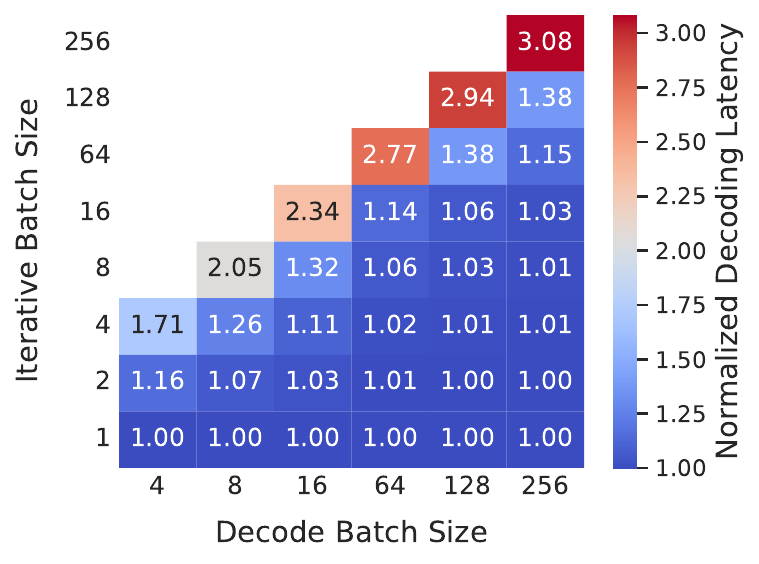}
    \caption{Performance degrade}
    \label{fig:case_3_batching_idleness_latency}
  \end{subfigure}
  \vspace{-1em}
  \caption{Decode idleness due to batched iterative queries.}
  \vspace{-.5em}
  \label{fig:case_3_batching}
\end{figure}

\Cref{fig:case_3_batching} further illustrates the phenomenon of decoding slowdown caused by idleness.
\Cref{fig:case_3_batching_idleness} visualizes the batching process, while the heatmap (\Cref{fig:case_3_batching_idleness_latency}) shows normalized decoding latency (compared to no retrieval) as a function of the decode batch size (x-axis) and iterative retrieval batch size (y-axis).
In this evaluation, the retrieval and prefix stages are assumed to have zero latency, isolating the slowdown to the batching-induced waiting time.
The results show that the effective latency is highly sensitive to the ratio of decode batch size to iterative retrieval batch size.
When these batch sizes are similar (\egc both set to 64), the normalized decoding latency reaches up to \xx{2.77$\times$}.
This increase occurs because one of the requests may generate a larger number of tokens before the next retrieval, resulting in idleness becomes a dominant factor.
For smaller ratios (e.g., decode batch size 64 and retrieval batch size up to 16), latency increases more gradually, indicating a more balanced workload with minimal idleness.
This observation aligns with \Cref{fig:case_3_latency_prefix_decode_batch}, where, for a decode batch size of 64, increasing the iterative retrieval batch size from 16 (\xx{1.14$\times$} normalized latency due to idleness) to 64 (\xx{2.77$\times$} normalized latency due to idleness) causes a significant increase in TPOT latency.
In summary, the results suggest that (a) when there is a large pool of XPUs that allows for large decoding batches, one can choose the iterative batch size that saturates database throughput, however, (b) with a smaller pool of XPUs and smaller decoding batch sizes, the optimal decoding batch size may actually be lower than the one that fully saturates the database.

\subsection{{\textup{\textbf{Case IV:}}} Query Rewriter and reranker}
\label{subsec:case4}
\begin{figure}[t]
  \centering
  \begin{subfigure}[b]{0.45\linewidth}
    \includegraphics[height=7em]
    {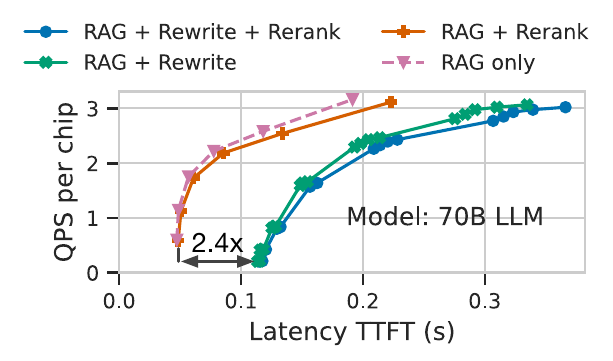}
  \end{subfigure}
  \begin{subfigure}[b]{0.54\linewidth}
    \includegraphics[height=7em]
    {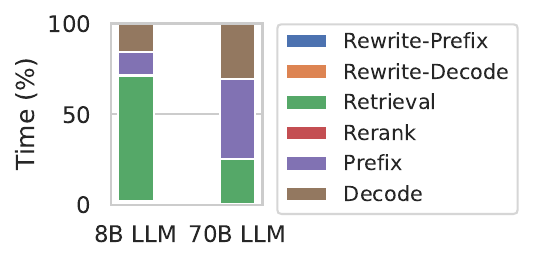}
  \end{subfigure}
  \vspace{-1.5em}
  \caption{RAG performance with rewriter and reranker. }
  \vspace{-1em}
  \label{fig:case_4_latency_throughput}
\end{figure}
In this setup, we extend Case I by integrating an 8B query rewriter model~\cite{dubey2024llama} and a 120M reranker~\cite{reimers2019sentence}.
The rewriter processes a 32-token question and generates a rephrased question of the same length, while the reranker evaluates 16 nearest passages, each containing 100 tokens, and returns the top five nearest neighbors.

\niparagraph{Takeaways:} 
While the reranker has negligible impact on overall RAG performance, the query rewriter can significantly increase TTFT latency due to its autoregressive nature.

\niparagraph{System performance with rewriter and reranker models.}
\Cref{fig:case_4_latency_throughput} (left) presents the performance of various RAG configurations with or without rewriter and reranker.
The results indicate that QPS/Chip remains largely unaffected by the addition of the rewriting and reranking modules.
This is further validated from  \Cref{fig:case_4_latency_throughput} which shows that negligible time is spent in rewriter and reranker stages.
However, the TTFT latency increases significantly (2.4$\times$) when the rewriter is included, due to its autoregressive generation nature, while reranking has minimal impact on TTFT.
This highlights the importance of considering an application’s latency tolerance when integrating the rewriter model.

\section{\oursemph: Systematic RAG Serving Optimization}
\label{sec:ragflow}
\begin{figure}[t]
  \centering
  \begin{subfigure}[b]{1.0\linewidth}
    \includegraphics[width=\linewidth]{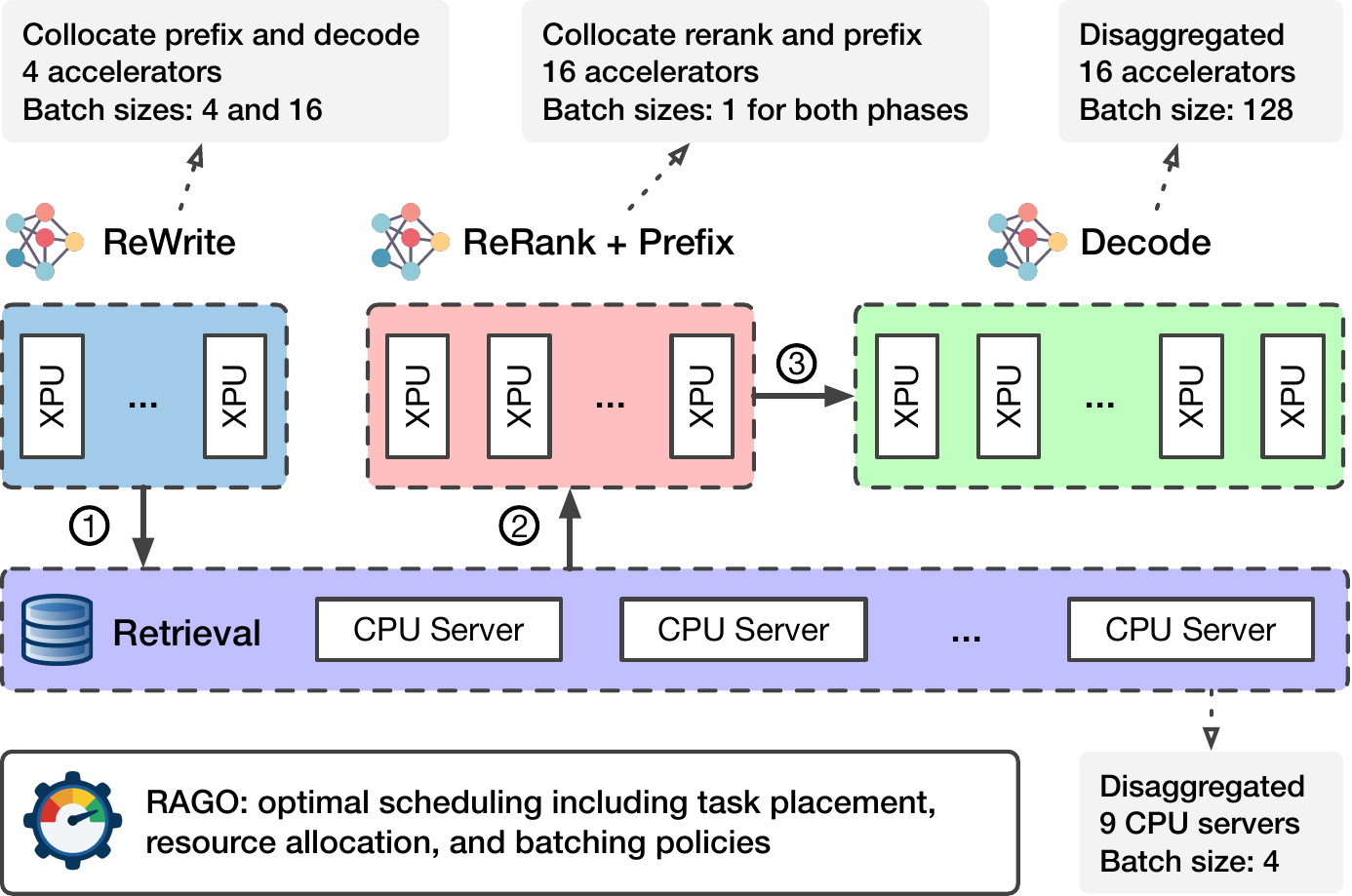}
  \end{subfigure}
  \vspace{-1em}
  \caption{An example of \ours optimizing placement, allocation, and batching policies for efficient RAG serving.}
  \vspace{-.5em}
  \label{fig:flexrag}
\end{figure}
Given the heterogeneous components and high workload variance across RAG (\S\ref{sec:case-studies}), one-size-fits-all systems are inherently inadequate for achieving optimal serving efficiency.
To overcome this challenge, we introduce \oursemph, a systematic framework to design and optimize RAG serving systems across diverse configurations.
\ours determines an optimized scheduling policy tailored to a specific \ragdesc and a defined performance target. 
The following sections expound on the components of \ours and its overall design.

\subsection{\ours Scheduling Decisions}
Each scheduling solution comprises three pivotal system decisions: \textit{task placement, resource allocation, and batching policy.}
\Cref{fig:flexrag} illustrates an example how these decisions come together to optimize a RAG serving pipeline under the constraint of 36 XPUs.
In this example, \ours adopts a hybrid collocation-disaggregation task placement strategy.
Specifically, the pipeline is organized into two collocated subsystems: (1) the rewrite-prefix and rewrite-decode phases; and (2) the rerank and prefix phases of response generation.
This organization ensures that tightly coupled tasks are efficiently grouped.
Resource allocation is tailored to the computational demands of each subsystem.
For instance, the query rewriter is assigned four XPUs, while the decoding phase, requiring significantly higher computational power, is allocated 16 XPUs.
To further enhance efficiency, \ours assigns batching policies customized to the characteristics of each phase.
For example, the rerank and prefix phases prioritize low-latency processing with a batch size of one, whereas the decoding phase operates with a much larger batch size of 128 to maximize throughput.
Below, we formally describe each system scheduling decision in \ours, deferring how to search for optimal schedules to~\S\ref{sec:rag_scheduler}.

\niparagraph{[I] Task placement.}
Recent LLM serving systems~\cite{patel2023splitwise,zhong2024distserve} advocate for disaggregating the prefix and decode phases (\S\ref{sec:background}), as these phases exhibit distinct workload characteristics\,---compute-bound vs. memory-bound---\,and impact TTFT versus TPOT.
However, given the multiple components in a RAG pipeline (\Cref{fig:rag-schema}), a natural question arises: \emph{should RAG systems adhere to the convention of fully disaggregated designs common in LLM-only systems?}

While prefix-decode disaggregation often proves beneficial (\S\ref{sec:background}), RAG pipelines may benefit more from collocation or hybrid strategies 
--- where some model components are collocated on the same set of XPUs while others are disaggregated over different XPUs ---\, particularly for components leading up to the prefix phase.
First, several components in the pipeline --- such as the database encoder, reranker, and the prefix phases of both the query rewriter and the main LLM --- share a similar profile of high computational intensity, and thus time-multiplexing these components on the same set of XPUs can inherently mitigate workload imbalances among them.
Second, components up to the prefix phase directly influence TTFT latency: while a fully disaggregated design, constrained by limited accelerator resources per stage, can prolong TTFT, collocation mitigates this by allowing all components to share the available resources, thereby reducing latency.

\begin{figure}[t]
  \centering
  \begin{subfigure}[b]{0.8\linewidth}
    \includegraphics[width=\linewidth]{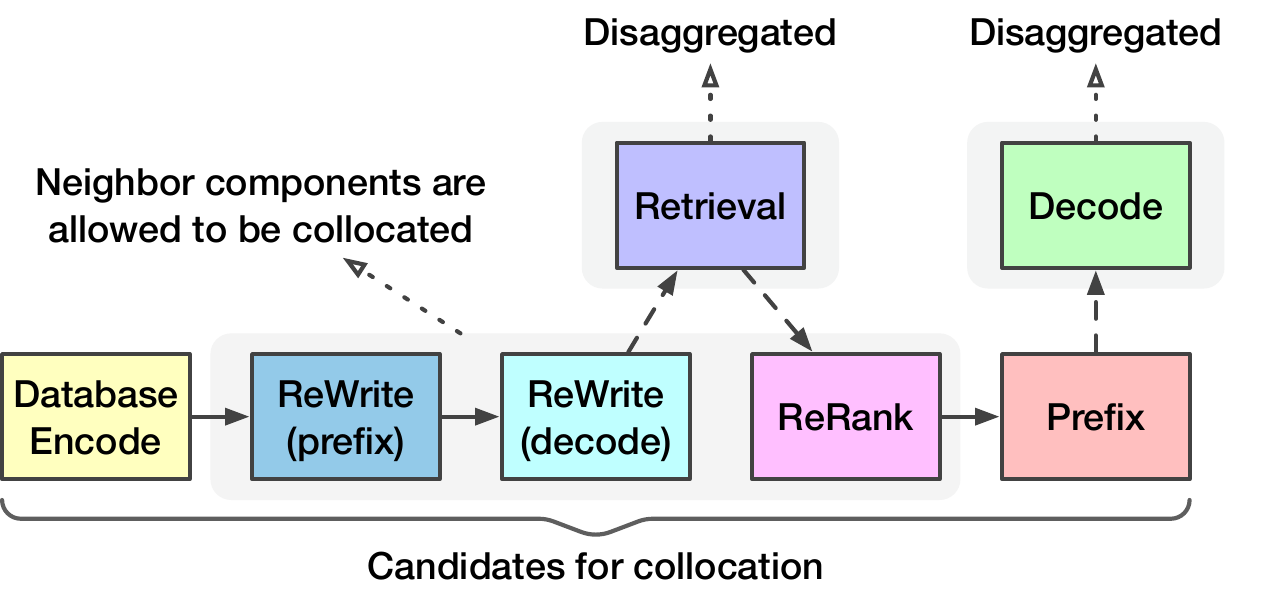}
  \end{subfigure}
  \vspace{-1em}
  \caption{\ours allows collocation of neighbor models.}
  \vspace{-.5em}
  \label{fig:collocation_example}
\end{figure}

That said, the decision between collocation and disaggregation depends on the specific characteristics of the RAG pipeline. 
For instance, the decoding phase of the query rewriter is autoregressive, and scales pooly with small batch sizes even with additional XPUs~\cite{vllm, yu2022orca}. 
Thus, collocating it with the prefix phase across many chips risks underutilizing hardware resources, as analyzed in~\S\ref{sec:eval}.
To address these challenges, \ours supports hybrid collocation-disaggregation task placement policies to balance flexibility and performance.
Specifically, we make the following assumptions regarding placement policies.
Firstly, the main LLM’s prefix and decode phases remain disaggregated, consistent with the strategies in~\cite{patel2023splitwise,zhong2024distserve}; 
Secondly, retrieval is always treated as a disaggregated task, as it operates on CPUs rather than XPUs.
Finally, neighboring phases up to the prefix can be collocated (\Cref{fig:collocation_example}). Collocation is restricted to consecutive neighbors to avoid excessively complicating the search space.

\niparagraph{[II] Resource allocation.}
After determining task placement, \ours assigns resources to each pipeline phase based on its computational and memory requirements.
For collocated inference phases, this involves selecting the appropriate number of accelerators to ensure efficient execution. 
Similarly, for retrieval operations, \ours determines the number of CPU servers required to meet workload demands.
The framework balances throughput requirements and latency constraints to ensure optimal performance.
Additionally, \ours ensures that each component has sufficient accelerator or CPU memory capacity to store the required models or database segments while meeting the specified performance targets.

\niparagraph{[III] Batching policy.}
Given a batch of incoming requests, \ours enables each stage of the pipeline to adopt varying batch sizes, offering flexibility to balance latency and throughput at each stage.
Upon receiving a burst of user requests, one can either use the same batch size for all stages before decoding or divide the requests into micro-batches with the same or different batch sizes.
For the decode stage, \ours leverage continuous batching~\cite{yu2022orca, vllm} to use larger batch sizes than individual requests, thereby improving throughput, as we evaluate in \S\ref{sec:eval}.
Moreover, in the case of iterative retrievals (\S\ref{subsec:case3}), \ours allows distinct batch sizes for the \emph{initial} retrieval/prefix pair and the \emph{subsequent} decoder-initiated retrieval/prefix iterations.
This differentiation is crucial because the initial retrieval and prefix phases directly affect TTFT, while the iterative ones primarily impact TPOT during token generation (see \S\ref{subsec:case3}).

\begin{figure}[t]
  \centering
  \begin{subfigure}[b]{0.8\linewidth}
    \includegraphics[width=\linewidth]{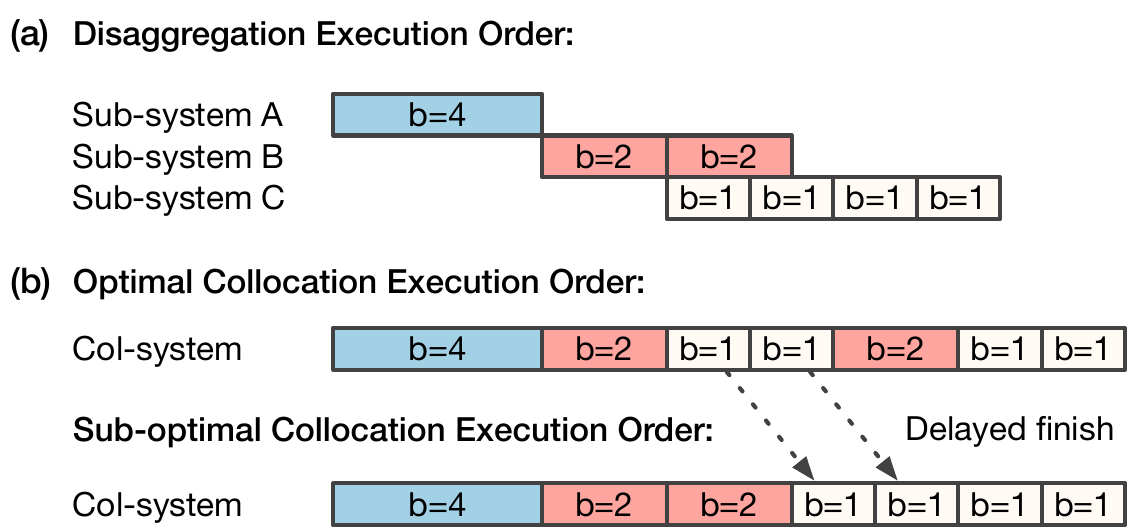}
  \end{subfigure}
  \vspace{-.5em}
  \caption{Execution order of batched requests until prefix.}
  \vspace{-.5em}
  \label{fig:execution_order}
\end{figure}
Once batch sizes are determined, \ours organizes their execution order to maximize efficiency based on the task placement strategy.
Here, we discuss the order of stages up to prefix, as the generative LLM decode always apply continuous batching~\cite{yu2022orca, vllm}.
In a fully disaggregated design (\Cref{fig:execution_order}(a)), execution is straightforward. As soon as (1) sufficient inputs arrive for a subsystem and (2) the subsystem completes its previous batch, it processes the new batch and forwards the output to the next subsystem.
On the other hand, \Cref{fig:execution_order}(b) shows the execution order of the collocated design. 
For simplicity, we use time-multiplexing strategy and leave more complex strategies such as simultaneous execution as future work.
In time-multiplexed designs, the throughput of the collocated system is fixed once batch sizes are set for each stage.
In such cases, a stage begins execution as soon as it accumulates enough inputs. 
As illustrated in \Cref{fig:execution_order}, the optimal execution order prioritizes completing the final stage (b=1) early over processing another round of the second stage (b=2), thereby minimizing the average completion time of the final stage.
If a retrieval operation is required between collocated stages (e.g., between the rewrite and prefix stages), the system pauses until the retrieval phase is complete before resuming the next collocated model inference phase.
\subsection{Searching for Optimal Scheduling Policies}
\label{sec:rag_scheduler}

Given a \ragdesc and hardware resource constraints, \ours performs an exhaustive search across potentially millions of schedules to identify Pareto frontier for key performance metrics such as TTFT and QPS/Chip.
Users can define the search granularity to constrain the search space, for example, by limiting the search to accelerator numbers or batch sizes that are powers of two.

As shown in Algorithm\ref{alg:search}, \ours search for optimal schedules in three main steps.
First, it performs performance profiling by evaluating each RAG component individually (\egc retrieval, prefix, etc.) under varying resource allocations and batch sizes.
This evaluation relies on the calibrated analytical model introduced in~\S\ref{sec:method}.
Next, it proceeds to schedule generation, where it explores \emph{all} possible RAG schedules by considering (a) collocation strategies, (b) resource allocation strategies within the overall resource budget, and (c) batching strategies for each component.
Finally, \ours calculates the end-to-end performance for each schedule and identifying the Pareto frontier along with the corresponding schedule configurations.

\begin{algorithm}[t]

\caption{Exhaustive Search for Optimal Scheduling}

\label{alg:search}

\DontPrintSemicolon
\SetKwInOut{Input}{Input}
\SetKwInOut{Output}{Output}

\Input{\( \ragdesc \) parameters, resource constraints \( RC \)}
\Output{Performance Pareto frontier and schedules \( P_{\text{RAG}} \)}

\textbf{Step 1: Performance Profiling Per RAG Stage} \;
Initialize \( P_{\text{stage}} = \emptyset \) \tcp*{Performance per stage}  

\ForEach{stage \( st \) in \( \ragdesc \)}{
    \( P_{\text{stage}}[st] = \emptyset \)  
    
    \ForEach{resource \( r \) where \( r < RC \)}{
        \( P_{\text{stage}}[st] \gets P_{\text{stage}}[st] \cup \text{Perf}(st, r) \) 
    }
    \( P_{\text{stage}}[st] \gets \text{getPareto}( P_{\text{stage}}[st]) \) 
}

\textbf{Step 2: Schedule Generation} \;
\( SC_{\text{p}} \gets \text{getPlacementOptions}(P_{\text{stage}}, RC) \) 

\( SC_{\text{a}} \gets \text{getAllocationOptions}(P_{\text{stage}}, RC, SC_{\text{p}}) \) 

\( SC_{\text{b}} \gets \text{getBatchingOptions}(P_{\text{stage}}, RC) \) 

\textbf{Step 3: End-to-end RAG Performance Evaluation} \;
Initialize \( P_{\text{RAG}} = \emptyset \) \tcp*{Performance of various schedules}  

\ForEach{schedule \( sc_{\text{e2e}} \in SC_{\text{p}} \times SC_{\text{a}} \times SC_{\text{b}} \)}{
    \( P_{\text{RAG}} \gets P_{\text{RAG}} \cup \text{assemblePerf}(sc_{\text{e2e}}, P_{\text{stage}}) \) 
}

\( P_{\text{RAG}} \gets \text{getPareto}(P_{\text{RAG}}) \) 

\Return \( P_{\text{RAG}} \) 
\tcp*{Pareto frontier and schedules}

\end{algorithm}

\section{Evaluation}%
\label{sec:eval}
We evaluate the effectiveness of \ours by revisiting the four RAG case studies in \S\ref{sec:case-studies}.
We begin with an analysis of the performance overview across all scheduling policies, followed by a detailed examination of each scheduling decision: placement, allocation, and batching.

\niparagraph{Evaluation methodology.}
For evaluating placement and resource allocation decisions, we focus on case study II (C-II)\,---long-context sequence---\,and case study IV (C-IV)\,---RAG with rewriter and reranker.
We select these case studies because of their additional model components, which distinguish them from LLM-only systems.
For micro-batching policy evaluations under bursts of user requests, we include case study I (C-I) with hyperscale retrieval, alongside C-II and C-IV. 
We exclude case study III (C-III), which focuses on iterative retrievals during decoding, as it was evaluated in details in~\S\ref{subsec:case3}.

\subsection{Overall Performance}
\label{sec:eval:e2e}

\niparagraph{Baseline system.}
Our baseline is an extension of LLM-only systems, where additional RAG components are collocated with the prefix system of the generative LLM.
Rather than arbitrarily assigning chips to prefix and decode, we carefully tune the ratio based on their time consumption. 
In this tuned baseline, the prefix and decode stages are allocated in a 1:1 chip ratio, reflecting their similar time requirements in the pipeline (1.2$\sim$1.4:1 across the 8B and 70B models). 
\begin{table*}[t]
    \centering
    \setlength\dashlinedash{0.2pt}
    \setlength\dashlinegap{1.5pt}
    \setlength\arrayrulewidth{0.3pt}
    \begin{footnotesize}
    \setlength{\tabcolsep}{4pt} %
    \renewcommand{\arraystretch}{0.85} %
    \caption{Comparison of \ours and baseline system schedules (placement, allocation, and batching strategies) in Case II.}
  \vspace{-1em}
    \label{tbl:rag-configurations}
    \scalebox{1.0}{ %
        \begin{tabular}{p{2.8cm}!{\vrule width 0.7pt}cc!{\vrule width 0.7pt}cccc!{\vrule width 0.7pt}cccc}
            \hline
            \multirow{2}{*}{\textbf{Schedules}} & \multicolumn{2}{c!{\vrule width 0.7pt}}{\textbf{Performance}} & \multicolumn{4}{c!{\vrule width 0.7pt}}{\textbf{Batch Sizes}} & \multicolumn{4}{c}{\textbf{Num XPUs}} \\ \cline{2-11}
             & \textbf{TTFT (s)} & \textbf{QPS/Chip} & \textbf{Encode} & \textbf{Retrieval} & \textbf{Prefix} & \textbf{Decode} & \textbf{Encode} & \textbf{Prefix} & \textbf{Decode} & \textbf{Total} \\ \hline
            RAGO (Max QPS/Chip)       & 2.47 & 1.08 & 2   & 2   & 128  & 1024 & 64 & 16 & 16 & 96  \\ \hdashline
            RAGO (Min TTFT)           & 0.03 & 0.22 & 1   & 1   & 1    & 128  & 64 (col) & 64 (col) & 64 & 128 \\ \hdashline
            Baseline (Max QPS/Chip)   & 1.54 & 0.65 & 2   & 2   & 128  & 256  & 64 (col) & 64 (col) & 64 & 128 \\ \hdashline
            Baseline (Min TTFT)       & 0.03 & 0.22 & 1   & 1   & 1    & 128  & 64 (col) & 64 (col) & 64 & 128 \\ \hline
        \end{tabular}
    } %
    \end{footnotesize}
\end{table*}

\begin{figure}[t]
  \centering
  \begin{subfigure}[b]{0.49\linewidth}
    \includegraphics[height=8em]
    {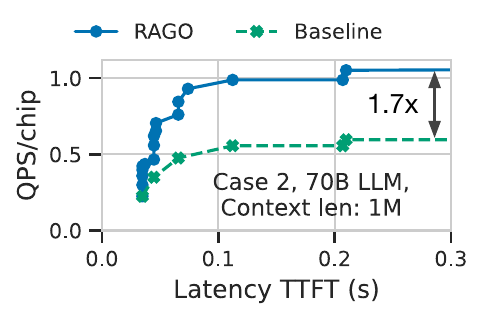}
    \caption{Case II}
    \label{fig:eval:e2e_case_2}
  \end{subfigure}
  \begin{subfigure}[b]{0.49\linewidth}
    \includegraphics[height=8em]
    {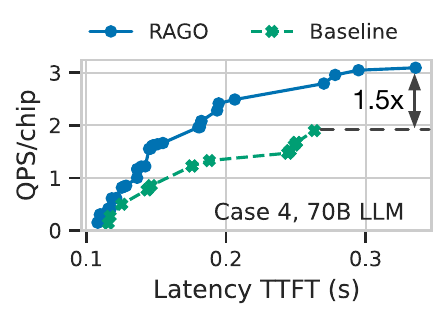}
    \caption{Case IV}
    \label{fig:eval:e2e_case_4}
  \end{subfigure}
  \vspace{-1em}
  \caption{\ours versus LLM-only system extension.}
  \vspace{-.5em}
  \label{fig:eval:e2e}
\end{figure}

\niparagraph{Impact of scheduling policies on QPS/Chip.}
\Cref{fig:eval:e2e_case_2} illustrates the Pareto performance comparison between \ours and the baseline in terms of QPS/Chip across two distinct RAG case studies.
In C-II, \ours achieves a \xx{1.7$\times$} improvement in maximum QPS/Chip over the baseline. 
This speedup underscores the inefficiencies of the baseline approach, particularly in handling the encoding stage for long-context sequences.
The encoder, while smaller than the generative LLM, becomes a critical bottleneck as context lengths grow. 
Specifically, in the baseline configuration, encoding is collocated with the prefix stage, leading to resource imbalance: decoding XPUs (\xx{50\%} of all XPUs) remain idle, while encode-prefix XPUs are overloaded. 
This imbalance can theoretically reduce QPS/Chip by up to \xx{2.0$\times$} in the baseline, which aligns with our observed reduction of \xx{1.94$\times$} for a large 10M-token context, though this specific data point is not plotted.
On the other hand, \ours achieves high QPS/Chip by allocating 64 out of the 96 XPUs to encoding (\Cref{tbl:rag-configurations}), reflecting the high time consumption of this stage. 

A similar inefficiency of the baseline is observed in C-IV (Figure~\ref{fig:eval:e2e_case_4}), where the rewriter and reranker models, despite their relatively small size (8B and 120M), significantly impact throughput in the baseline system. 
This QPS drop can be attributed to two primary factors.
First, collocating rewriter-decode stage and the prefix stage of the main generative LLM leads to XPU under-utilization due to the low computational intensity of the autoregressive decoding stage, particularly when handling small batch sizes.
Second, retrieval operations are introduced between the rewriting and prefix stages add wait times for retrieval results
(e.g., 10 ms with a batch size of one given 32 host servers), further reducing throughput.
In contrast, \ours demonstrates its ability to mitigate these bottlenecks through optimized task placement, resource allocation, and batching strategies. 
These results highlight the importance of disaggregating smaller pipeline stages and balancing resource distribution to unlock the full throughput potential of RAG systems.

\begin{figure}[t]
  \centering
  \begin{subfigure}[b]{0.49\linewidth}
    \includegraphics[height=7em]
    {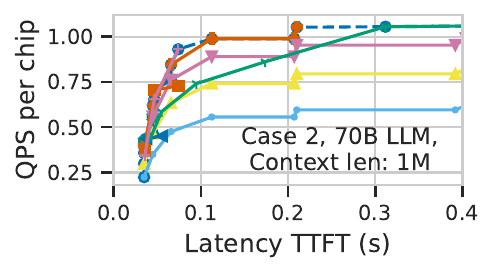}
    \caption{Case II}
    \label{fig:eval:pareto_breakdown_case_2}
  \end{subfigure}
  \begin{subfigure}[b]{0.49\linewidth}
    \includegraphics[height=7em]
    {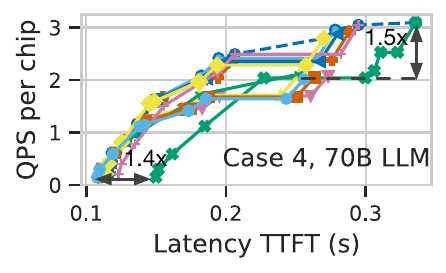}
    \caption{Case IV}
    \label{fig:eval:pareto_breakdown_case_4}
  \end{subfigure}
  \vspace{-1em}
  \caption{Performance Pareto across multiple placement and allocation plans in case 2 and 4.}
  \vspace{-.5em}
  \label{fig:eval:pareto_breakdown}
\end{figure}

\niparagraph{Pareto composition analysis.}
\Cref{fig:eval:pareto_breakdown_case_2} and \Cref{fig:eval:pareto_breakdown_case_4} reveal how diverse placement and allocation plans contribute to the overall Pareto frontier.
The dashed lines represent the global Pareto frontier, while each solid line corresponds to the Pareto frontier of a specific combination of placement and allocation strategies, with each point on a line representing a batching policy. 
We omit the legend for each plan, as providing detailed descriptions for each would take up excessive space, but will explain some of these plans below.
The overall Pareto frontier is constructed from multiple distinct plans, each embodying a unique trade-off between TTFT and QPS/Chip.%
This diversity underscores the importance of tailoring placement and allocation strategies to the specific performance priorities of a deployment.
For instance, as shown in Figure \ref{fig:eval:pareto_breakdown_case_4}, selecting the most throughput-optimized plan results in a trade-off, with TTFT approximately \xx{40\%} higher compared to the most latency-optimized plan, while achieving 1.5$\times$ QPS/Chip.
This is because the throughput-optimized plan allocates only one chip to the query rewriter, given its minimal contribution to the end-to-end generation latency, as analyzed in \S\ref{subsec:case4}.
In contrast, the latency-optimized plan allocates 32 chips to the query rewriter, resulting in low resource utilization since a significant number of chips are assigned to this non-bottleneck stage.
These findings emphasize that there is no one-size-fits-all strategy. 
Instead, the optimal placement and allocation plans must be aligned with the operational objectives, whether minimizing latency, maximizing throughput, or achieving a balance between the two.

\subsection{Scheduling Policy Sensitivity Analysis}
\label{sec:eval:schedule}

We now delve into a detailed analysis of the performance implication of each scheduling decision.

\niparagraph{Task placement sensitivity.}
\begin{figure}[t]
  \centering
  \begin{subfigure}[b]{0.45\linewidth}
    \includegraphics[height=8em]
    {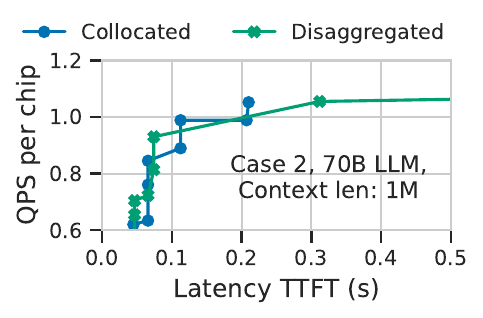}
    \caption{Case II}
    \label{fig:eval:placement_case_2}
  \end{subfigure}
  \begin{subfigure}[b]{0.54\linewidth}
    \includegraphics[height=8em]
    {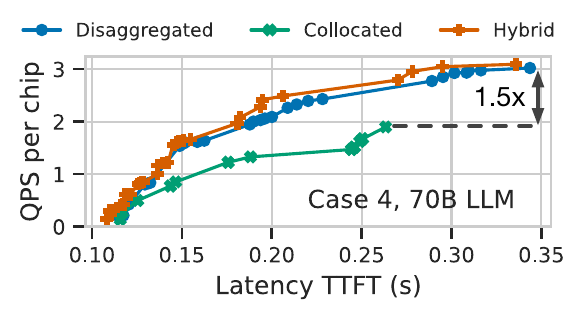}
    \caption{Case IV}
    \label{fig:eval:placement_case_4}
  \end{subfigure}
  \centering
    \vspace{-2em}
  \caption{Comparison of task placement plans.}
    \vspace{-.5em}
  \label{fig:eval:placement}
\end{figure}
\Cref{fig:eval:placement} compares the impact of different task placement policies on system performance across C-II and C-IV.
Each line in \Cref{fig:eval:placement_case_2} and \Cref{fig:eval:placement_case_4} represents the Pareto frontier for a specific placement strategy, illustrating the relationship between QPS/Chip and TTFT latency under these policies.

In C-II (\Cref{fig:eval:placement_case_2}), task placement sensitivity is minimal.
Both collocated and disaggregated strategies yield comparable performance, as the encoder and prefix stages are computationally intensive. 
Whether these stages are time-multiplexed (collocated) or spatially multiplexed (disaggregated), performance remains consistent (only 2\% difference in max QPS/Chip) as long as the accelerator ratio between stages is appropriately chosen.
This demonstrates that task placement decisions in this case have little effect on system efficiency, provided resources are balanced effectively.

In contrast, C-IV (\Cref{fig:eval:placement_case_4}) shows a more pronounced sensitivity to placement policies. 
Here, a hybrid placement strategy\,---where some model components are collocated on the same set of XPUs while others are disaggregated over different XPUs ---\, slightly outperforms the fully disaggregated approach and significantly surpasses the collocated plan, achieving up to a \xx{1.5$\times$} improvement in QPS/Chip. 
The key advantage of the hybrid and disaggregated strategies lies in their ability to mitigate the underutilization of prefix chips, which occurs when the rewriter model is collocated with the prefix stage. 
By separating the rewriter model from the prefix system, these strategies prevent resource bottlenecks and enable optimal throughput.

\begin{figure}[t]
  \centering
  \begin{subfigure}[b]{0.49\linewidth}
    \includegraphics[width=\linewidth]{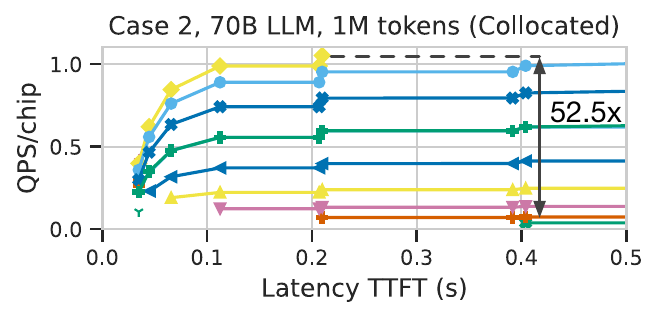}
    \caption{Collocated}
    \label{fig:eval:resource_case_2_collocated}
  \end{subfigure}
  \begin{subfigure}[b]{0.49\linewidth}
    \includegraphics[width=\linewidth]{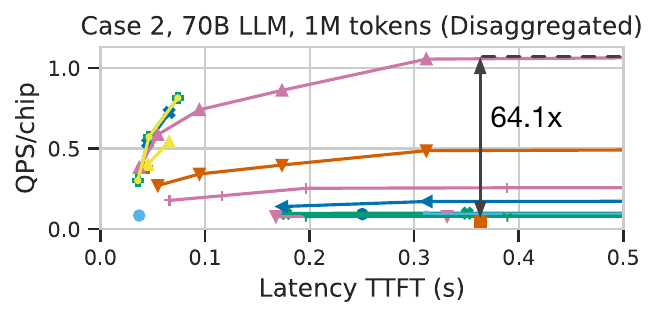}
    \caption{Disaggregated}
    \label{fig:eval:resource_case_2_disaggregated}
  \end{subfigure}
  \centering
  \vspace{-1em}
  \caption{Comparison of resource allocation plans (case II).}
  \vspace{-.5em}
  \label{fig:eval:resource_case_2}
\end{figure}
\niparagraph{Resource allocation sensitivity.}
\Cref{fig:eval:resource_case_2} shows the high sensitivity of performance to resource allocation strategies regardless of whether collocated or disaggregated placement plans are used.
Each curve represents the Pareto frontier of a specific resource allocation strategy in Case II.
Due to space constraints, the legend for each plan is omitted, with explanations for selected plans provided below.
For collocated plans, the maximum QPS/chip can vary by up to 52.5$\times$ if insufficient resources are allocated to high-workload stages, when other stages have surplus capacity. 
For example, in the collocated plan (\Cref{fig:eval:resource_case_2_collocated}), imbalanced resource distribution across the encoder and prefix stages leads to underutilization of available accelerators, limiting throughput.
This effect may amplify to 64.1$\times$ QPS/chip difference for disaggregated plans, as disaggregated stages rely heavily on precise balancing to maximize performance. 
Tailoring resource distribution to the specific demands of each stage is essential for optimizing both latency and throughput in RAG systems.

\begin{figure}[t]
  \centering
  \begin{subfigure}[b]{0.32\linewidth}
    \includegraphics[height=6em]
    {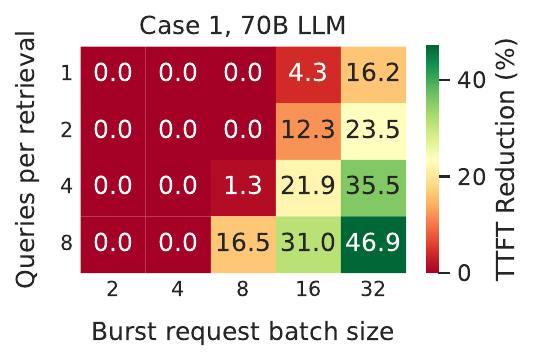}
    \caption{Case I}
    \label{fig:eval:microbatch_case_1}
  \end{subfigure}
  \begin{subfigure}[b]{0.35\linewidth}
    \includegraphics[height=6em]
    {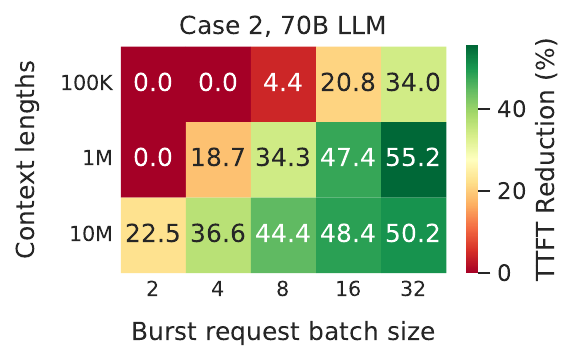}
    \caption{Case II}
    \label{fig:eval:microbatch_case_2}
  \end{subfigure}
  \begin{subfigure}[b]{0.3\linewidth}
    \includegraphics[height=6em]
    {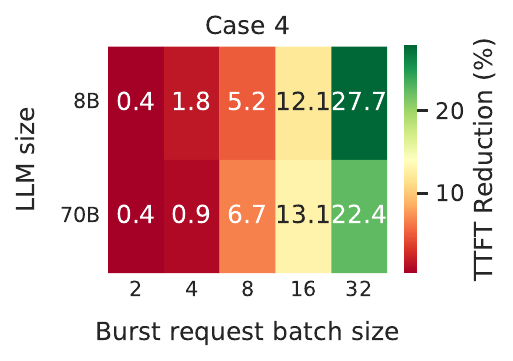}
    \caption{Case IV}
    \label{fig:eval:microbatch_case_4}
  \end{subfigure}
  \vspace{-1em}
  \caption{TTFT latency reduction by micro-batching.}
  \vspace{-.5em}
  \label{fig:eval:microbatch}
\end{figure}

\niparagraph{Impact of micro-batching on TTFT latency.} 
The effectiveness of micro-batching depends on the throughput sensitivity to batch size at each pipeline stage.
When a burst of user requests arrives, they can be divided into micro-batches across the stages prior to decoding.
\Cref{fig:eval:microbatch} compares the impact of micro-batching on TTFT reduction across the case studies. 
For C-II (\Cref{fig:eval:microbatch_case_2}), micro-batching is effective even with a small batch size of two, reducing TTFT by 22\%. 
This is because both the encoding and prefix stages are computationally intensive, achieving reasonable throughput even with smaller batch sizes. 
With larger batches of 32, the TTFT reduction increases further to 55\% for 1M tokens.
In C-I (\Cref{fig:eval:microbatch_case_1}), micro-batching only becomes effective with larger batch sizes, such as eight or 16.
This inefficiency at smaller batch sizes arises from the vector search system, where reducing the query batch size below 16 fails to improve latency.
However, with batch sizes increasing to 32, micro-batching still achieves a significant latency reduction of 46\% for eight queries per vector.
For C-IV (\Cref{fig:eval:microbatch_case_4}), TTFT reduction is moderate, with a maximum improvement of approximately \xx{25\%} at a batch size of 32. 
This modest improvement is primarily due to the query rewriter decoding stage, which exhibits little latency reduction with smaller sequence batch sizes.

\section{Additional Related Work}
\label{sec:related_work}

\niparagraph{RAG performance optimization.} 
As an emerging research area, RAG performance optimization remains underexplored, with existing studies targeting specific configurations.
Chameleon\cite{jiang2023chameleon} integrates both retrieval and LLM accelerators within RAG systems. Its retrieval accelerator is tailored for large-scale, product-quantization-based vector search, making it particularly beneficial in hyper-scale retrieval scenarios, where retrieval constitutes a major bottleneck~(\S\ref{subsec:case1}).
PipeRAG\cite{jiang2024piperag} and RaLMSpec\cite{zhang2024accelerating} address decoding stalls in iterative retrievals (\S\ref{subsec:case3}) through data prefetching --- PipeRAG employs approximate prefetching, whereas RaLMSpec incorporates an additional asynchronous verification step.
Additionally, since RAG introduces increased prompt length and associated computation costs (\S\ref{subsec:case1}), studies such as CacheBlend\cite{yao2024cacheblend} and RAGCache~\cite{jin2024ragcache} propose caching the KV-cache of database documents, a method that is particularly effective when the prompt-to-generation cost ratio is high.
If these techniques are adopted, the workload distribution within a RAG system evaluated by \ours is expected to shift.
For example, retrieval acceleration~\cite{jiang2023chameleon} will shift the workload toward being more inference-bound; pre-computing KV cache of retrieved documents~\cite{yao2024cacheblend, jin2024ragcache} will increase the importance of retrieval and decoding performance; supporting data prefetching in iterative retrievals through ~\cite{jiang2024piperag, zhang2024accelerating} will reduce decoding engine idleness during retrieval operations.

\niparagraph{LLM and retrieval optimization.}  
Extensive research has been devoted to optimizing LLM systems and their underlying hardware~\cite{patel2023splitwise, zhong2024distserve, vllm, yu2022orca, qin2024mecla, zhang2024llmcompass, zhao2024alisa, lee2024tender, li2024large, bang2023vtrain, yun2024duplex, shao2019simba}. 
Similarly, significant efforts have focused on enhancing retrieval efficiency, a fundamental topic in databases~\cite{adb-v, divya_rdbms, mailthody2019deepstore}, which is essential not only for RAG but also for recommender systems~\cite{google_recommendation, suchal2010full}, scientific applications~\cite{bajusz2015tanimoto, woodbridge2016improving}, and video matching~\cite{garcia2008fast, shao2008batch, adb-v}.  
These efforts include both algorithm- and system-level optimizations~\cite{malkov2018efficient, PQ, chen2021spann, jayaram2019diskann, zhang2023vbase, xu2023spfresh, mohoney2023high, pan2023survey} as well as modern hardware supports for various retrieval algorithms, spanning product-quantization-based ANN algorithms~\cite{jiang2023co, johnson2019billion, lee2022anna, liu2023juno} and graph-based approaches~\cite{jiang2024accelerating, zeng2023df, groh2022ggnn, zhao2020song, jang2023cxl}.  
However, the complexity and heterogeneity of RAG pipelines far exceed those of LLM-only or retrieval-only systems, highlighting the need for \ours to perform system-level RAG scheduling.

\section{Conclusion and Future Outlook}
\label{sec:conclusion}

This work represents an early exploration of RAG through a systems lens, establishing a foundation for this rapidly evolving field.
Our contribution, \ragdesc, provides a structured abstraction for RAG serving, facilitating systematic workload characterization and bridging the gap between algorithms and system design.
Leveraging \ragdesc, we proposed \ours, a system optimization framework that delivers up to a 2$\times$ improvement in QPS per chip and a 55\% reduction in TTFT compared to a strong baseline.

As the field advances, our characterization results can guide the design of future RAG systems and  hardware. 
For instance, our findings highlight that retrieval can become a bottleneck in certain RAG paradigms, particularly at hyperscale, underscoring the need for further retrieval optimizations.
Moreover, with multiple model components in RAG pipelines, efficient support for collocated models on accelerators will be increasingly critical.
Finally, scaling to extremely complicated RAG and agentic AI systems introduces challenges related to a broader optimization search space and additional efficiency metrics, such as energy and cost efficiency. 
We leave these investigations to future work.

\section*{Acknowledgements}
We extend our gratitude towards Cliff Young, David Culler, and Eugene Le for reviewing the paper and providing insightful feedback.
We also thank the extended team at Google DeepMind and System Research@Google who enabled and supported this research direction.

\bibliographystyle{ACM-Reference-Format}
\bibliography{refs}

\begin{thebibliography}{100}
\providecommand{\url}[1]{#1}
\csname url@samestyle\endcsname
\providecommand{\newblock}{\relax}
\providecommand{\bibinfo}[2]{#2}
\providecommand{\BIBentrySTDinterwordspacing}{\spaceskip=0pt\relax}
\providecommand{\BIBentryALTinterwordstretchfactor}{4}
\providecommand{\BIBentryALTinterwordspacing}{\spaceskip=\fontdimen2\font plus
\BIBentryALTinterwordstretchfactor\fontdimen3\font minus \fontdimen4\font\relax}
\providecommand{\BIBforeignlanguage}[2]{{%
\expandafter\ifx\csname l@#1\endcsname\relax
\typeout{** WARNING: IEEEtran.bst: No hyphenation pattern has been}%
\typeout{** loaded for the language `#1'. Using the pattern for}%
\typeout{** the default language instead.}%
\else
\language=\csname l@#1\endcsname
\fi
#2}}
\providecommand{\BIBdecl}{\relax}
\BIBdecl

\bibitem{brown2020language}
T.~Brown, B.~Mann, N.~Ryder, M.~Subbiah, J.~D. Kaplan, P.~Dhariwal, A.~Neelakantan, P.~Shyam, G.~Sastry, A.~Askell \emph{et~al.}, ``{Language Models are Few-shot Learners},'' in \emph{NeurIPS}, 2020.

\bibitem{chowdhery2022palm}
A.~Chowdhery, S.~Narang, J.~Devlin, M.~Bosma, G.~Mishra, A.~Roberts, P.~Barham, H.~W. Chung, C.~Sutton, S.~Gehrmann \emph{et~al.}, ``{PaLM: Scaling Language Modeling with Pathways},'' \emph{arXiv preprint arXiv:2204.02311}, 2022.

\bibitem{dubey2024llama}
A.~Dubey, A.~Jauhri, A.~Pandey, A.~Kadian, A.~Al-Dahle, A.~Letman, A.~Mathur, A.~Schelten, A.~Yang, A.~Fan \emph{et~al.}, ``{The Llama 3 Herd of Models},'' \emph{arXiv preprint arXiv:2407.21783}, 2024.

\bibitem{li2022competition}
Y.~Li, D.~Choi, J.~Chung, N.~Kushman, J.~Schrittwieser, R.~Leblond, T.~Eccles, J.~Keeling, F.~Gimeno, A.~Dal~Lago \emph{et~al.}, ``{Competition-Level Code Generation with AlphaCode},'' \emph{Science}, 2022.

\bibitem{liu2024your}
J.~Liu, C.~S. Xia, Y.~Wang, and L.~Zhang, ``{Is your Code Generated by ChatGPT Really Correct? Rigorous Evaluation of Large Language Models for Code Generation},'' in \emph{NeurIPS}, 2024.

\bibitem{poesia2022synchromesh}
G.~Poesia, O.~Polozov, V.~Le, A.~Tiwari, G.~Soares, C.~Meek, and S.~Gulwani, ``{Synchromesh: Reliable Code Generation From Pre-trained Language Models},'' \emph{arXiv preprint arXiv:2201.11227}, 2022.

\bibitem{beltagy2019scibert}
I.~Beltagy, K.~Lo, and A.~Cohan, ``{SciBERT: A Pretrained Language Model for Scientific Text},'' \emph{arXiv preprint arXiv:1903.10676}, 2019.

\bibitem{taylor2022galactica}
R.~Taylor, M.~Kardas, G.~Cucurull, T.~Scialom, A.~Hartshorn, E.~Saravia, A.~Poulton, V.~Kerkez, and R.~Stojnic, ``{Galactica: A Large Language Model for Science},'' \emph{arXiv preprint arXiv:2211.09085}, 2022.

\bibitem{lee2020biobert}
J.~Lee, W.~Yoon, S.~Kim, D.~Kim, S.~Kim, C.~H. So, and J.~Kang, ``{BioBERT: A Pre-trained Biomedical Language Representation Model for Biomedical Text Mining},'' \emph{Bioinformatics}, 2020.

\bibitem{lazaridou2021mind}
A.~Lazaridou, A.~Kuncoro, E.~Gribovskaya, D.~Agrawal, A.~Liska, T.~Terzi, M.~Gimenez, C.~de~Masson~d'Autume, T.~Kocisky, S.~Ruder \emph{et~al.}, ``{Mind the Gap: Assessing Temporal Generalization in Neural Language Models},'' in \emph{NeurIPS}, 2021.

\bibitem{lewis2020retrieval}
P.~Lewis, E.~Perez, A.~Piktus, F.~Petroni, V.~Karpukhin, N.~Goyal, H.~K{\"u}ttler, M.~Lewis, W.-t. Yih, T.~Rockt{\"a}schel \emph{et~al.}, ``{Retrieval-Augmented Generation for Knowledge-Intensive NLP Tasks},'' in \emph{NeurIPS}, 2020.

\bibitem{li2023dark}
Z.~Li, ``{The Dark Side of ChatGPT: Legal and Ethical Challenges From Stochastic Parrots and Hallucination},'' \emph{arXiv preprint arXiv:2304.14347}, 2023.

\bibitem{ji2023survey}
Z.~Ji, N.~Lee, R.~Frieske, T.~Yu, D.~Su, Y.~Xu, E.~Ishii, Y.~J. Bang, A.~Madotto, and P.~Fung, ``{Survey of Hallucination in Natural Language Generation},'' \emph{ACM Computing Surveys}, 2023.

\bibitem{izacard2020leveraging}
G.~Izacard and E.~Grave, ``{Leveraging Passage Retrieval with Generative Models for Open Domain Question Answering},'' \emph{arXiv preprint arXiv:2007.01282}, 2020.

\bibitem{fan2019eli5}
A.~Fan, Y.~Jernite, E.~Perez, D.~Grangier, J.~Weston, and M.~Auli, ``{ELI5: Long form Question Answering},'' \emph{arXiv preprint arXiv:1907.09190}, 2019.

\bibitem{kasneci2023chatgpt}
E.~Kasneci, K.~Se{\ss}ler, S.~K{\"u}chemann, M.~Bannert, D.~Dementieva, F.~Fischer, U.~Gasser, G.~Groh, S.~G{\"u}nnemann, E.~H{\"u}llermeier \emph{et~al.}, ``{ChatGPT for Good? On Opportunities and Challenges of Large Language Models for Education},'' \emph{Learning and Individual Differences}, 2023.

\bibitem{thirunavukarasu2023large}
A.~J. Thirunavukarasu, D.~S.~J. Ting, K.~Elangovan, L.~Gutierrez, T.~F. Tan, and D.~S.~W. Ting, ``{Large language Models in Medicine},'' \emph{Nature medicine}, 2023.

\bibitem{borgeaud2022improving}
S.~Borgeaud, A.~Mensch, J.~Hoffmann, T.~Cai, E.~Rutherford, K.~Millican, G.~B. Van Den~Driessche, J.-B. Lespiau, B.~Damoc, A.~Clark \emph{et~al.}, ``{Improving Language Models by Retrieving From Trillions of Tokens},'' in \emph{ICML}, 2022.

\bibitem{shao2024scaling}
R.~Shao, J.~He, A.~Asai, W.~Shi, T.~Dettmers, S.~Min, L.~Zettlemoyer, and P.~W. Koh, ``{Scaling Retrieval-Based Language Models with a Trillion-Token Datastore},'' \emph{arXiv preprint arXiv:2407.12854}, 2024.

\bibitem{wang2023instructretro}
B.~Wang, W.~Ping, L.~McAfee, P.~Xu, B.~Li, M.~Shoeybi, and B.~Catanzaro, ``{InstructRetro: Instruction Tuning post Retrieval-Augmented Pretraining},'' \emph{arXiv preprint arXiv:2310.07713}, 2023.

\bibitem{lewis2020pre}
M.~Lewis, M.~Ghazvininejad, G.~Ghosh, A.~Aghajanyan, S.~Wang, and L.~Zettlemoyer, ``{Pre-training via Paraphrasing},'' in \emph{NeurIPS}, 2020.

\bibitem{guu2020realm}
K.~Guu, K.~Lee, Z.~Tung, P.~Pasupat, and M.-W. Chang, ``{REALM: Retrieval-Augmented Language Model Pre-Training},'' \emph{arXiv preprint arXiv:2002.08909}, 2020.

\bibitem{edge2024local}
D.~Edge, H.~Trinh, N.~Cheng, J.~Bradley, A.~Chao, A.~Mody, S.~Truitt, and J.~Larson, ``{From Local to Global: A Graph RAG Approach to Query-Focused Summarization},'' \emph{arXiv preprint arXiv:2404.16130}, 2024.

\bibitem{ruminer2024notebooklm}
M.~Ruminer, ``{Google's NotebookLM and RAG},'' 2024.

\bibitem{databricksrag}
Databricks, ``{RAG (Retrieval Augmented Generation) on Databricks},'' 2024.

\bibitem{llamaindex}
Meta, ``{Build AI Knowledge Assistants over your Enterprise Data},'' 2024.

\bibitem{team2024gemini}
G.~Team, P.~Georgiev, V.~I. Lei, R.~Burnell, L.~Bai, A.~Gulati, G.~Tanzer, D.~Vincent, Z.~Pan, S.~Wang \emph{et~al.}, ``{Gemini 1.5: Unlocking Multimodal Understanding Across Millions of Tokens of Context},'' \emph{arXiv preprint arXiv:2403.05530}, 2024.

\bibitem{reimers2019sentence}
N.~Reimers and I.~Gurevych, ``{Sentence-BERT: Sentence Embeddings using Siamese BERT-Networks},'' \emph{arXiv preprint arXiv:1908.10084}, 2019.

\bibitem{lee2024gecko}
J.~Lee, Z.~Dai, X.~Ren, B.~Chen, D.~Cer, J.~R. Cole, K.~Hui, M.~Boratko, R.~Kapadia, W.~Ding \emph{et~al.}, ``{Gecko: Versatile Text Embeddings Distilled From Large Language Models},'' \emph{arXiv preprint arXiv:2403.20327}, 2024.

\bibitem{chan2024rq}
C.-M. Chan, C.~Xu, R.~Yuan, H.~Luo, W.~Xue, Y.~Guo, and J.~Fu, ``{RQ-RAG: Learning to Refine Queries for Retrieval Augmented Generation},'' \emph{arXiv preprint arXiv:2404.00610}, 2024.

\bibitem{ma2023query}
X.~Ma, Y.~Gong, P.~He, H.~Zhao, and N.~Duan, ``{Query Rewriting for Retrieval-Augmented Large Language Models},'' \emph{arXiv preprint arXiv:2305.14283}, 2023.

\bibitem{glass2022re2g}
M.~Glass, G.~Rossiello, M.~F.~M. Chowdhury, A.~R. Naik, P.~Cai, and A.~Gliozzo, ``{Re2G: Retrieve, Rerank, Generate},'' \emph{arXiv preprint arXiv:2207.06300}, 2022.

\bibitem{allahverdiyev2024chunkrag}
R.~A. I. S. S.~I. Allahverdiyev, M.~Taha, A.~Akalin, and K.~Zhu, ``{ChunkRAG: Novel LLM-Chunk Filtering Method for RAG Systems},'' \emph{arXiv preprint arXiv:2410.19572}, 2024.

\bibitem{lee2024can}
J.~Lee, A.~Chen, Z.~Dai, D.~Dua, D.~S. Sachan, M.~Boratko, Y.~Luan, S.~M. Arnold, V.~Perot, S.~Dalmia \emph{et~al.}, ``{Can Long-Context Language Models Subsume Retrieval, RAG, SQL, and More?}'' \emph{arXiv preprint arXiv:2406.13121}, 2024.

\bibitem{li2024retrieval}
Z.~Li, C.~Li, M.~Zhang, Q.~Mei, and M.~Bendersky, ``{Retrieval Augmented Generation or Long-context LLMs? A Comprehensive Study and Hybrid Approach},'' \emph{arXiv preprint arXiv:2407.16833}, 2024.

\bibitem{yue2024inference}
Z.~Yue, H.~Zhuang, A.~Bai, K.~Hui, R.~Jagerman, H.~Zeng, Z.~Qin, D.~Wang, X.~Wang, and M.~Bendersky, ``{Inference Scaling for Long-Context Retrieval Augmented Generation},'' \emph{arXiv preprint arXiv:2410.04343}, 2024.

\bibitem{trivedi2022interleaving}
H.~Trivedi, N.~Balasubramanian, T.~Khot, and A.~Sabharwal, ``{Interleaving Retrieval with Chain-of-Thought Reasoning for Knowledge-Intensive Multi-Step Questions},'' \emph{arXiv preprint arXiv:2212.10509}, 2022.

\bibitem{jiang2023active}
Z.~Jiang, F.~F. Xu, L.~Gao, Z.~Sun, Q.~Liu, J.~Dwivedi-Yu, Y.~Yang, J.~Callan, and G.~Neubig, ``{Active Retrieval Augmented Generation},'' \emph{arXiv preprint arXiv:2305.06983}, 2023.

\bibitem{parashar2019timeloop}
A.~Parashar, P.~Raina, Y.~S. Shao, Y.-H. Chen, V.~A. Ying, A.~Mukkara, R.~Venkatesan, B.~Khailany, S.~W. Keckler, and J.~Emer, ``{Timeloop: A Systematic Approach to DNN Accelerator Evaluation},'' in \emph{ISPASS}, 2019.

\bibitem{zhang2022full}
D.~Zhang, S.~Huda, E.~Songhori, K.~Prabhu, Q.~Le, A.~Goldie, and A.~Mirhoseini, ``{A Full-Stack Search Technique for Domain Optimized Deep Learning Accelerators},'' in \emph{ASPLOS}, 2022.

\bibitem{huang2024mind}
Q.~Huang, P.-A. Tsai, J.~S. Emer, and A.~Parashar, ``{Mind the Gap: Attainable Data Movement and Operational Intensity Bounds for Tensor Algorithms},'' in \emph{ISCA}, 2024.

\bibitem{jouppi2017datacenter}
N.~P. Jouppi, C.~Young, N.~Patil, D.~Patterson, G.~Agrawal, R.~Bajwa, S.~Bates, S.~Bhatia, N.~Boden, A.~Borchers \emph{et~al.}, ``{In-Datacenter Performance Analysis of a Tensor Processing Unit},'' in \emph{ISCA}, 2017.

\bibitem{inferentia}
``{AWS Inferentia},'' \url{https://aws.amazon.com/ai/machine-learning/inferentia/}.

\bibitem{xu2023nearest}
F.~F. Xu, U.~Alon, and G.~Neubig, ``{Why do Nearest Neighbor Language Models Work?}'' \emph{arXiv preprint arXiv:2301.02828}, 2023.

\bibitem{yu2024rankrag}
Y.~Yu, W.~Ping, Z.~Liu, B.~Wang, J.~You, C.~Zhang, M.~Shoeybi, and B.~Catanzaro, ``{RankRAG: Unifying Context Ranking with Retrieval-Augmented Generation in LLMs},'' \emph{arXiv preprint arXiv:2407.02485}, 2024.

\bibitem{yang2024crag}
X.~Yang, K.~Sun, H.~Xin, Y.~Sun, N.~Bhalla, X.~Chen, S.~Choudhary, R.~D. Gui, Z.~W. Jiang, Z.~Jiang \emph{et~al.}, ``{CRAG--Comprehensive RAG Benchmark},'' \emph{arXiv preprint arXiv:2406.04744}, 2024.

\bibitem{ram2023context}
O.~Ram, Y.~Levine, I.~Dalmedigos, D.~Muhlgay, A.~Shashua, K.~Leyton-Brown, and Y.~Shoham, ``{In-Context Retrieval-Augmented Language Models},'' \emph{arXiv preprint arXiv:2302.00083}, 2023.

\bibitem{komeili2021internet}
M.~Komeili, K.~Shuster, and J.~Weston, ``{Internet-Augmented Dialogue Generation},'' \emph{arXiv preprint arXiv:2107.07566}, 2021.

\bibitem{guu2020retrieval}
K.~Guu, K.~Lee, Z.~Tung, P.~Pasupat, and M.~Chang, ``{Retrieval Augmented Language Model Pre-training},'' in \emph{ICML}, 2020.

\bibitem{khandelwal2019generalization}
U.~Khandelwal, O.~Levy, D.~Jurafsky, L.~Zettlemoyer, and M.~Lewis, ``{Generalization through Memorization: Nearest Neighbor Language Models},'' \emph{arXiv preprint arXiv:1911.00172}, 2019.

\bibitem{khandelwal2020nearest}
U.~Khandelwal, A.~Fan, D.~Jurafsky, L.~Zettlemoyer, and M.~Lewis, ``{Nearest Neighbor Machine Translation},'' \emph{arXiv preprint arXiv:2010.00710}, 2020.

\bibitem{smith2022using}
S.~Smith, M.~Patwary, B.~Norick, P.~LeGresley, S.~Rajbhandari, J.~Casper, Z.~Liu, S.~Prabhumoye, G.~Zerveas, V.~Korthikanti \emph{et~al.}, ``{Using DeepSpeed and Megatron to Train Megatron-Turing NLG 530B, A Large-Scale Generative Language Model},'' \emph{arXiv preprint arXiv:2201.11990}, 2022.

\bibitem{rae2021scaling}
J.~W. Rae, S.~Borgeaud, T.~Cai, K.~Millican, J.~Hoffmann, F.~Song, J.~Aslanides, S.~Henderson, R.~Ring, S.~Young \emph{et~al.}, ``{Scaling Language Models: Methods, Analysis \& Insights From Training Gopher},'' \emph{arXiv preprint arXiv:2112.11446}, 2021.

\bibitem{patel2023splitwise}
P.~Patel, E.~Choukse, C.~Zhang, {\'I}.~Goiri, A.~Shah, S.~Maleki, and R.~Bianchini, ``{Splitwise: Efficient Generative LLM Inference Using Phase Splitting},'' \emph{arXiv preprint arXiv:2311.18677}, 2023.

\bibitem{zhong2024distserve}
Y.~Zhong, S.~Liu, J.~Chen, J.~Hu, Y.~Zhu, X.~Liu, X.~Jin, and H.~Zhang, ``{DistServe: Disaggregating Prefill and Decoding for Goodput-optimized Large Language Model Serving},'' in \emph{OSDI}, 2024.

\bibitem{vaswani2017attention}
A.~Vaswani, N.~Shazeer, N.~Parmar, J.~Uszkoreit, L.~Jones, A.~N. Gomez, {\L}.~Kaiser, and I.~Polosukhin, ``{Attention Is All You Need},'' in \emph{NeurIPS}, 2017.

\bibitem{ke2022disaggrec}
L.~Ke, X.~Zhang, B.~Lee, G.~E. Suh, and H.-H.~S. Lee, ``{DisaggRec: Architecting Disaggregated Systems for Large-Scale Personalized Recommendation},'' \emph{arXiv preprint arXiv:2212.00939}, 2022.

\bibitem{jiang2021fleetrec}
W.~Jiang, Z.~He, S.~Zhang, K.~Zeng, L.~Feng, J.~Zhang, T.~Liu, Y.~Li, J.~Zhou, C.~Zhang \emph{et~al.}, ``{FleetRec: Large-scale Recommendation Inference on Hybrid GPU-FPGA Clusters},'' in \emph{SIGKDD}, 2021.

\bibitem{PQ}
H.~Jegou, M.~Douze, and C.~Schmid, ``{Product Quantization for Nearest Neighbor Search},'' \emph{IEEE Transactions on Pattern Analysis and Machine Intelligence}, 2010.

\bibitem{malkov2018efficient}
Y.~A. Malkov and D.~A. Yashunin, ``{Efficient and Robust Approximate Nearest Neighbor Search using Hierarchical Navigable Small World Graphs},'' \emph{IEEE Transactions on Pattern Analysis and Machine Intelligence}, 2018.

\bibitem{malkov2014approximate}
Y.~Malkov, A.~Ponomarenko, A.~Logvinov, and V.~Krylov, ``{Approximate Nearest Neighbor Algorithm based on Navigable Small World Graphs},'' \emph{Information Systems}, 2014.

\bibitem{fu2017fast}
C.~Fu, C.~Xiang, C.~Wang, and D.~Cai, ``{Fast Approximate Nearest Neighbor Search with the Navigating Spreading-out Graph},'' \emph{arXiv preprint arXiv:1707.00143}, 2017.

\bibitem{zhao2023towards}
X.~Zhao, Y.~Tian, K.~Huang, B.~Zheng, and X.~Zhou, ``{Towards Efficient Index Construction and Approximate Nearest Neighbor Search in High-Dimensional Spaces},'' \emph{VLDB}, 2023.

\bibitem{zuo2023arkgraph}
C.~Zuo and D.~Deng, ``{ARKGraph: All-Range Approximate K-Nearest-Neighbor Graph},'' \emph{VLDB}, 2023.

\bibitem{lu2021hvs}
K.~Lu, M.~Kudo, C.~Xiao, and Y.~Ishikawa, ``{HVS: Hierarchical Graph Structure based on Voronoi Diagrams for Solving Approximate Nearest Neighbor Search},'' \emph{VLDB}, 2021.

\bibitem{gao2023high}
J.~Gao and C.~Long, ``{High-dimensional Approximate Nearest Neighbor Search: With Reliable and Efficient Distance Comparison Operations},'' \emph{Proceedings of the ACM on Management of Data}, 2023.

\bibitem{johnson2019billion}
J.~Johnson, M.~Douze, and H.~J{\'e}gou, ``{Billion-Scale Similarity Search with GPUs},'' \emph{IEEE Transactions on Big Data}, 2019.

\bibitem{jiang2023chameleon}
W.~Jiang, M.~Zeller, R.~Waleffe, T.~Hoefler, and G.~Alonso, ``{Chameleon: A Heterogeneous and Disaggregated Accelerator System for Retrieval-Augmented Language Models},'' \emph{VLDB}, 2025.

\bibitem{faiss}
``{Faiss},'' \url{https://github.com/facebookresearch/faiss /}.

\bibitem{guo2020accelerating}
R.~Guo, P.~Sun, E.~Lindgren, Q.~Geng, D.~Simcha, F.~Chern, and S.~Kumar, ``{Accelerating Large-Scale Inference with Anisotropic Vector Quantization},'' in \emph{ICML}, 2020.

\bibitem{scann_github}
``{ScaNN: Scalable Nearest Neighbors},'' \url{https://github.com/google-research/google-research/blob/master/scann}.

\bibitem{andre2016cache}
F.~Andr{\'e}, A.-M. Kermarrec, and N.~Le~Scouarnec, ``{Cache Locality is not Enough: High-performance Nearest Neighbor Search with Product Quantization Fast Scan},'' in \emph{VLDB}, 2016.

\bibitem{sun2024soar}
P.~Sun, D.~Simcha, D.~Dopson, R.~Guo, and S.~Kumar, ``{SOAR: Improved Indexing for Approximate Nearest Neighbor Search},'' in \emph{NeurIPS}, 2024.

\bibitem{notebooklm}
``{NotebookLM: Note Taking and Research Assistant Powered by AI},'' \url{ https://notebooklm.google/}.

\bibitem{chatgpt}
``{OpenAI ChatGPT},'' \url{https://chat.openai.com/}.

\bibitem{rag_github}
``{Advanced RAG Techniques: Elevating Your Retrieval-Augmented Generation Systems},'' \url{ https://github.com/NirDiamant/RAG_Techniques}.

\bibitem{wang2024richrag}
S.~Wang, X.~Xu, M.~Wang, W.~Chen, Y.~Zhu, and Z.~Dou, ``{RichRAG: Crafting Rich Responses for Multi-faceted Queries in Retrieval-Augmented Generation},'' \emph{arXiv preprint arXiv:2406.12566}, 2024.

\bibitem{besta2024multi}
M.~Besta, A.~Kubicek, R.~Niggli, R.~Gerstenberger, L.~Weitzendorf, M.~Chi, P.~Iff, J.~Gajda, P.~Nyczyk, J.~M{\"u}ller \emph{et~al.}, ``{Multi-Head RAG: Solving Multi-Aspect Problems with LLMs},'' \emph{arXiv preprint arXiv:2406.05085}, 2024.

\bibitem{annbench}
``{ANN-Benchmarks: A Benchmarking Environment for Approximate Nearest Neighbor Algorithms Search},'' \url{ https://ann-benchmarks.com/}.

\bibitem{SIFT}
\BIBentryALTinterwordspacing
``{SIFT ANNS dataset}.'' [Online]. Available: \url{http://corpus-texmex.irisa.fr/}
\BIBentrySTDinterwordspacing

\bibitem{babenko2016efficient}
A.~Babenko and V.~Lempitsky, ``{Efficient Indexing of Billion-Scale Datasets of Deep Descriptors},'' in \emph{CVPR}, 2016.

\bibitem{simhadri2022results}
H.~V. Simhadri, G.~Williams, M.~Aum{\"u}ller, M.~Douze, A.~Babenko, D.~Baranchuk, Q.~Chen, L.~Hosseini, R.~Krishnaswamy, G.~Srinivasa \emph{et~al.}, ``{Results of the NeurIPS'21 Challenge on Billion-Scale Approximate Nearest Neighbor Search},'' \emph{arXiv preprint arXiv:2205.03763}, 2022.

\bibitem{sun2023automating}
P.~Sun, R.~Guo, and S.~Kumar, ``{Automating Nearest Neighbor Search Configuration with Constrained Optimization},'' \emph{arXiv preprint arXiv:2301.01702}, 2023.

\bibitem{bajaj2016ms}
P.~Bajaj, D.~Campos, N.~Craswell, L.~Deng, J.~Gao, X.~Liu, R.~Majumder, A.~McNamara, B.~Mitra, T.~Nguyen \emph{et~al.}, ``{MS MARCO: A Human Generated MAchine Reading COmprehension Dataset},'' \emph{arXiv preprint arXiv:1611.09268}, 2016.

\bibitem{joshi2017triviaqa}
M.~Joshi, E.~Choi, D.~S. Weld, and L.~Zettlemoyer, ``{TriviaQA: A Large Scale Distantly Supervised Challenge Dataset for Reading Comprehension},'' \emph{arXiv preprint arXiv:1705.03551}, 2017.

\bibitem{rajpurkar2018know}
P.~Rajpurkar, R.~Jia, and P.~Liang, ``{Know What You Don't Know: Unanswerable Questions for SQuAD},'' \emph{arXiv preprint arXiv:1806.03822}, 2018.

\bibitem{asai2023self}
A.~Asai, Z.~Wu, Y.~Wang, A.~Sil, and H.~Hajishirzi, ``{Self-RAG: Learning to Retrieve, Generate, and Critique through Self-Reflection},'' \emph{arXiv preprint arXiv:2310.11511}, 2023.

\bibitem{sharegpt}
``{ShareGPT: Share your ChatGPT conversations},'' \url{ https://sharegpt.com/ }.

\bibitem{vllm}
W.~Kwon, Z.~Li, S.~Zhuang, Y.~Sheng, L.~Zheng, C.~H. Yu, J.~E. Gonzalez, H.~Zhang, and I.~Stoica, ``{Efficient Memory Management for Large Language Model Serving with PagedAttention},'' in \emph{SIGOPS}, 2023.

\bibitem{tpuv5e}
``{TPU v5e},'' \url{ https://cloud.google.com/tpu/docs/v5e }, 2023.

\bibitem{tpuv4}
``{TPU v4},'' \url{https://cloud.google.com/tpu/docs/v4}, 2021.

\bibitem{tpuv5p}
``{TPU v5p},'' \url{ https://cloud.google.com/tpu/docs/v5p }, 2023.

\bibitem{huang2019gpipe}
Y.~Huang, Y.~Cheng, A.~Bapna, O.~Firat, D.~Chen, M.~Chen, H.~Lee, J.~Ngiam, Q.~V. Le, Y.~Wu \emph{et~al.}, ``{GPipe: Efficient Training of Giant Neural Networks using Pipeline Parallelism},'' in \emph{NeurIPS}, 2019.

\bibitem{narayanan2019pipedream}
D.~Narayanan, A.~Harlap, A.~Phanishayee, V.~Seshadri, N.~R. Devanur, G.~R. Ganger, P.~B. Gibbons, and M.~Zaharia, ``{PipeDream: Generalized Pipeline Parallelism for DNN Training},'' in \emph{SOSP}, 2019.

\bibitem{shoeybi2019megatron}
M.~Shoeybi, M.~Patwary, R.~Puri, P.~LeGresley, J.~Casper, and B.~Catanzaro, ``{Megatron-LM: Training Multi-Billion Parameter Language Models Using Model Parallelism},'' \emph{arXiv preprint arXiv:1909.08053}, 2019.

\bibitem{rajbhandari2020zero}
S.~Rajbhandari, J.~Rasley, O.~Ruwase, and Y.~He, ``{ZeRO: Memory Optimizations Toward Training Trillion Parameter Models},'' in \emph{SC}, 2020.

\bibitem{yu2022orca}
G.-I. Yu, J.~S. Jeong, G.-W. Kim, S.~Kim, and B.-G. Chun, ``{Orca: A Distributed Serving System for Transformer-Based Generative Models},'' in \emph{OSDI}, 2022.

\bibitem{jiang2024piperag}
W.~Jiang, S.~Zhang, B.~Han, J.~Wang, B.~Wang, and T.~Kraska, ``{PipeRAG: Fast Retrieval-Augmented Generation via Algorithm-System Co-design},'' in \emph{KDD}, 2025.

\bibitem{2024ragannquality}
A.~Leto, C.~Aguerrebere, I.~Bhati, T.~Willke, M.~Tepper, and V.~A. Vo, ``{Toward Optimal Search and Retrieval for RAG},'' \emph{arXiv preprint arXiv:2411.07396}, 2024.

\bibitem{zhang2024accelerating}
Z.~Zhang, A.~Zhu, L.~Yang, Y.~Xu, L.~Li, P.~M. Phothilimthana, and Z.~Jia, ``{Accelerating Retrieval-Augmented Language Model Serving with Speculation},'' \emph{arXiv preprint arXiv:2401.14021}, 2024.

\bibitem{yao2024cacheblend}
J.~Yao, H.~Li, Y.~Liu, S.~Ray, Y.~Cheng, Q.~Zhang, K.~Du, S.~Lu, and J.~Jiang, ``{CacheBlend: Fast Large Language Model Serving with Cached Knowledge Fusion},'' \emph{arXiv preprint arXiv:2405.16444}, 2024.

\bibitem{jin2024ragcache}
C.~Jin, Z.~Zhang, X.~Jiang, F.~Liu, X.~Liu, X.~Liu, and X.~Jin, ``{RAGCache: Efficient Knowledge Caching for Retrieval-Augmented Generation},'' \emph{arXiv preprint arXiv:2404.12457}, 2024.

\bibitem{qin2024mecla}
Y.~Qin, Y.~Wang, Z.~Zhao, X.~Yang, Y.~Zhou, S.~Wei, Y.~Hu, and S.~Yin, ``{MECLA: Memory-Compute-Efficient LLM Accelerator with Scaling Sub-matrix Partition},'' in \emph{ISCA}, 2024.

\bibitem{zhang2024llmcompass}
H.~Zhang, A.~Ning, R.~B. Prabhakar, and D.~Wentzlaff, ``{LLMCompass: Enabling Efficient Hardware Design for Large Language Model Inference},'' in \emph{ISCA}, 2024.

\bibitem{zhao2024alisa}
Y.~Zhao, D.~Wu, and J.~Wang, ``{ALISA: Accelerating Large Language Model Inference via Sparsity-Aware KV Caching},'' \emph{arXiv preprint arXiv:2403.17312}, 2024.

\bibitem{lee2024tender}
J.~Lee, W.~Lee, and J.~Sim, ``{Tender: Accelerating Large Language Models via Tensor Decomposition and Runtime Requantization},'' \emph{arXiv preprint arXiv:2406.12930}, 2024.

\bibitem{li2024large}
J.~Li, J.~Xu, S.~Huang, Y.~Chen, W.~Li, J.~Liu, Y.~Lian, J.~Pan, L.~Ding, H.~Zhou \emph{et~al.}, ``{Large Language Model Inference Acceleration: A Comprehensive Hardware Perspective},'' \emph{arXiv preprint arXiv:2410.04466}, 2024.

\bibitem{bang2023vtrain}
J.~Bang, Y.~Choi, M.~Kim, Y.~Kim, and M.~Rhu, ``{vTrain: A Simulation Framework for Evaluating Cost-effective and Compute-optimal Large Language Model Training},'' \emph{arXiv preprint arXiv:2312.12391}, 2023.

\bibitem{yun2024duplex}
S.~Yun, K.~Kyung, J.~Cho, J.~Choi, J.~Kim, B.~Kim, S.~Lee, K.~Sohn, and J.~H. Ahn, ``{Duplex: A Device for Large Language Models with Mixture of Experts, Grouped Query Attention, and Continuous Batching},'' \emph{arXiv preprint arXiv:2409.01141}, 2024.

\bibitem{shao2019simba}
Y.~S. Shao, J.~Clemons, R.~Venkatesan, B.~Zimmer, M.~Fojtik, N.~Jiang, B.~Keller, A.~Klinefelter, N.~Pinckney, P.~Raina \emph{et~al.}, ``{Simba: Scaling Deep-Learning Inference with Multi-Chip-Module-Based Architecture},'' in \emph{MICRO}, 2019.

\bibitem{adb-v}
C.~Wei, B.~Wu, S.~Wang, R.~Lou, C.~Zhan, F.~Li, and Y.~Cai, ``Analyticdb-v: a hybrid analytical engine towards query fusion for structured and unstructured data,'' \emph{Proceedings of the VLDB Endowment}, vol.~13, no.~12, pp. 3152--3165, 2020.

\bibitem{divya_rdbms}
D.~Mahajan, J.~K. Kim, J.~Sacks, A.~Ardalan, A.~Kumar, and H.~Esmaeilzadeh, ``{In-RDBMS Hardware Acceleration of Advanced Analytics},'' \emph{VLDB}, 2018.

\bibitem{mailthody2019deepstore}
V.~S. Mailthody, Z.~Qureshi, W.~Liang, Z.~Feng, S.~G. De~Gonzalo, Y.~Li, H.~Franke, J.~Xiong, J.~Huang, and W.-m. Hwu, ``{DeepStore: In-Storage Acceleration for Intelligent Queries},'' in \emph{MICRO}, 2019.

\bibitem{google_recommendation}
P.~Covington, J.~Adams, and E.~Sargin, ``{Deep Neural Networks for YouTube Recommendations},'' in \emph{RecSys}, 2016.

\bibitem{suchal2010full}
J.~Suchal and P.~N{\'a}vrat, ``{Full Text Search Engine as Scalable K-Nearest Neighbor Recommendation System},'' in \emph{IFIP}, 2010.

\bibitem{bajusz2015tanimoto}
D.~Bajusz, A.~R{\'a}cz, and K.~H{\'e}berger, ``{Why is Tanimoto Index an Appropriate Choice for Fingerprint-Based Similarity Calculations?}'' \emph{Journal of Cheminformatics}, pp. 1--13, 2015.

\bibitem{woodbridge2016improving}
J.~Woodbridge, B.~Mortazavi, A.~A. Bui, and M.~Sarrafzadeh, ``{Improving Biomedical Signal Search Results in Big Data Case-Based Reasoning Environments},'' \emph{Pervasive and mobile computing}, 2016.

\bibitem{garcia2008fast}
V.~Garcia, E.~Debreuve, and M.~Barlaud, ``{Fast K Nearest Neighbor Search using GPU},'' in \emph{IEEE Computer Society Conference on Computer Vision and Pattern Recognition Workshops}, 2008.

\bibitem{shao2008batch}
J.~Shao, Z.~Huang, H.~T. Shen, X.~Zhou, E.-P. Lim, and Y.~Li, ``{Batch Nearest Neighbor Search for Video Retrieval},'' \emph{IEEE Transactions on Multimedia}, 2008.

\bibitem{chen2021spann}
Q.~Chen, B.~Zhao, H.~Wang, M.~Li, C.~Liu, Z.~Li, M.~Yang, and J.~Wang, ``{SPANN: Highly-efficient Billion-scale Approximate Nearest Neighbor Search},'' \emph{arXiv preprint arXiv:2111.08566}, 2021.

\bibitem{jayaram2019diskann}
S.~Jayaram~Subramanya, F.~Devvrit, H.~V. Simhadri, R.~Krishnawamy, and R.~Kadekodi, ``{DiskANN: Fast Accurate Billion-point Nearest Neighbor Search on a Single Node},'' 2019.

\bibitem{zhang2023vbase}
Q.~Zhang, S.~Xu, Q.~Chen, G.~Sui, J.~Xie, Z.~Cai, Y.~Chen, Y.~He, Y.~Yang, F.~Yang \emph{et~al.}, ``{VBASE: Unifying Online Vector Similarity Search and Relational Queries via Relaxed Monotonicity},'' in \emph{OSDI}, 2023.

\bibitem{xu2023spfresh}
Y.~Xu, H.~Liang, J.~Li, S.~Xu, Q.~Chen, Q.~Zhang, C.~Li, Z.~Yang, F.~Yang, Y.~Yang \emph{et~al.}, ``{SPFresh: Incremental In-Place Update for Billion-Scale Vector Search},'' in \emph{SOSP}, 2023.

\bibitem{mohoney2023high}
J.~Mohoney, A.~Pacaci, S.~R. Chowdhury, A.~Mousavi, I.~F. Ilyas, U.~F. Minhas, J.~Pound, and T.~Rekatsinas, ``{High-Throughput Vector Similarity Search in Knowledge Graphs},'' \emph{Proceedings of the ACM on Management of Data}, 2023.

\bibitem{pan2023survey}
J.~J. Pan, J.~Wang, and G.~Li, ``{Survey of Vector Database Management Systems},'' \emph{arXiv preprint arXiv:2310.14021}, 2023.

\bibitem{jiang2023co}
W.~Jiang, S.~Li, Y.~Zhu, J.~de~Fine~Licht, Z.~He, R.~Shi, C.~Renggli, S.~Zhang, T.~Rekatsinas, T.~Hoefler, and G.~Alonso, ``{Co-design Hardware and Algorithm for Vector Search},'' in \emph{SC}, 2023.

\bibitem{lee2022anna}
Y.~Lee, H.~Choi, S.~Min, H.~Lee, S.~Beak, D.~Jeong, J.~W. Lee, and T.~J. Ham, ``{ANNA: Specialized Architecture for Approximate Nearest Neighbor Search},'' in \emph{HPCA}, 2022.

\bibitem{liu2023juno}
Z.~Liu, W.~Ni, J.~Leng, Y.~Feng, C.~Guo, Q.~Chen, C.~Li, M.~Guo, and Y.~Zhu, ``{JUNO: Optimizing High-Dimensional Approximate Nearest Neighbour Search with Sparsity-Aware Algorithm and Ray-Tracing Core Mapping},'' \emph{arXiv preprint arXiv:2312.01712}, 2023.

\bibitem{jiang2024accelerating}
W.~Jiang, H.~Hu, T.~Hoefler, and G.~Alonso, ``{Accelerating Graph-based Vector Search via Delayed-Synchronization Traversal},'' \emph{arXiv preprint arXiv:2406.12385}, 2024.

\bibitem{zeng2023df}
S.~Zeng, Z.~Zhu, J.~Liu, H.~Zhang, G.~Dai, Z.~Zhou, S.~Li, X.~Ning, Y.~Xie, H.~Yang \emph{et~al.}, ``{DF-GAS: a Distributed FPGA-as-a-Service Architecture towards Billion-Scale Graph-based Approximate Nearest Neighbor Search},'' in \emph{MICRO}, 2023.

\bibitem{groh2022ggnn}
F.~Groh, L.~Ruppert, P.~Wieschollek, and H.~P. Lensch, ``{GGNN: Graph-Based GPU Nearest Neighbor Search},'' \emph{IEEE Transactions on Big Data}, 2022.

\bibitem{zhao2020song}
W.~Zhao, S.~Tan, and P.~Li, ``{SONG: Approximate Nearest Neighbor Search on GPU},'' in \emph{ICDE}, 2020.

\bibitem{jang2023cxl}
J.~Jang, H.~Choi, H.~Bae, S.~Lee, M.~Kwon, and M.~Jung, ``{CXL-ANNS: Software-Hardware Collaborative Memory Disaggregation and Computation for Billion-Scale Approximate Nearest Neighbor Search},'' in \emph{ATC}, 2023.

\end{thebibliography}
\end{document}